 \documentclass{aa}

\usepackage{graphicx}
\usepackage{txfonts}
\usepackage[dvipsnames,table]{xcolor}
\usepackage[utf8]{inputenc}
\usepackage{xspace}
\usepackage{natbib}
\usepackage{pgfplots}
\graphicspath{{gfx/}}

\bibpunct{(}{)}{;}{a}{}{,}
\usepackage{hyperref}
\hypersetup{colorlinks=true, linktocpage=true, pdfstartpage=3,
            pdfstartview=FitV, breaklinks=true, pdfpagemode=UseNone,
            pageanchor=true, pdfpagemode=UseOutlines, plainpages=false,
            bookmarksnumbered, bookmarksopen=true, bookmarksopenlevel=1,
            hypertexnames=true, pdfhighlight=/O, urlcolor=RoyalBlue,
            linkcolor=RoyalBlue, citecolor=RoyalBlue}

\usepackage{catoptions}
\makeatletter

\def\Autoref#1{%
  \begingroup
  \edef\reserved@a{\cpttrimspaces{#1}}%
  \ifcsndefTF{r@#1}{%
    \xaftercsname{\expandafter\testreftype\@fourthoffive}
      {r@\reserved@a}.\\{#1}%
  }{%
    \ref{#1}%
  }%
  \endgroup
}
\def\testreftype#1.#2\\#3{%
  \ifcsndefTF{#1autorefname}{%
    \def\reserved@a##1##2\@nil{%
      \uppercase{\def\ref@name{##1}}%
      \csn@edef{#1autorefname}{\ref@name##2}%
      \autoref{#3}%
    }%
    \reserved@a#1\@nil
  }{%
    \autoref{#3}%
  }%
}
\makeatother

\newcommand{\source}[1]{\textsuperscript{\textcolor{blue}{[citation needed]}}\xspace}

\newcommand{\skybot}{\texttt{SkyBoT}\xspace}
\newcommand{\sm}{\texttt{SkyMapper}\xspace}

\newcommand{\numb}[1]{\textcolor{orange}{#1}}
\renewcommand{\numb}[1]{#1}  

\renewcommand{\arcsec}{\ensuremath{^{\prime\prime}}\xspace}

\DeclareUnicodeCharacter{2212}{-}

\begin{document}
\newcommand\mpcmoc{550,000\xspace}   
\newcommand\mpcnow{960,000\xspace}   
\newcommand\nummoc{471,569\xspace}    
\newcommand\nummocid{211,138\xspace}  
\newcommand\numsvoiduniq{36,730\xspace} 
\newcommand\nummociduniq{100,322\xspace} 
\newcommand\numsvoid{57,646\xspace}  

\newcommand\initsrc{43,501,635\xspace}   
\newcommand\onccd{16,915,791\xspace}     
\newcommand\skmsrc{9,367,950\xspace}   
\newcommand\skmmatch{2,047,588\xspace}   
\newcommand\skmtot{880,528\xspace}   
\newcommand\skmuniq{205,515\xspace}   

\newcommand\skmgaia{8,203,916\xspace}  

\newcommand\colorsmes{669,545\xspace}   
\newcommand\colorsindmes{375,160\xspace} 

\newcommand\colorcalc{13,783\xspace}   
\newcommand\colortot{683,328\xspace}   
\newcommand\colorsindsso{388,943\xspace}   
\newcommand\indssowithcolors{139,220\xspace} 


\newcommand\taxgrgiiz{32,776\xspace} 
\newcommand\taxgriz{39,063\xspace} 
\newcommand\taxgiiz{33,299\xspace} 
\newcommand\taxriiz{35,498\xspace} 
\newcommand\taxgr{117,356\xspace} 

  \title{Multi-filter photometry of Solar System Objects from the SkyMapper Southern Survey
    \thanks{The catalogs presented here are available
    at the CDS via anonymous ftp to
    \url{http://cdsarc.u-strasbg.fr/} or via
    \url{http://cdsarc.u-strasbg.fr/viz-bin/qcat?J/A+A/xxx/Axxx}}}

  \author{A. V. Sergeyev\inst{\ref{i:oca},\ref{i:kha}} \and
          B. Carry\inst{\ref{i:oca}} \and
          C. A. Onken\inst{\ref{i:anu},\ref{i:cga}} \and
          H. A. R. Devillepoix\inst{\ref{i:curtin}} \and
          C. Wolf\inst{\ref{i:anu},\ref{i:cga}} \and
          S.-W. Chang\inst{\ref{i:anu},\ref{i:snu1},\ref{i:snu2}} 
          }
  \institute{Université Côte d'Azur, Observatoire de la Côte d'Azur, CNRS, Laboratoire Lagrange, France\\
	   \email{alexey.sergeyev@oca.eu; benoit.carry@oca.eu}
	   \label{i:oca} 
       \and
       Research School of Astronomy and Astrophysics, Australian National University, Canberra, ACT 2611, Australia\label{i:anu}
       \and 
       Centre for Gravitational Astrophysics, College of Science, The Australian National University, ACT 2601, Australia\label{i:cga} 
       \and
       School of Earth and Planetary Sciences, Curtin University, Perth WA 6845, Australia\label{i:curtin}
       \and SNU Astronomy Research Center, Seoul National University, 1 Gwanak-rho, Gwanak-gu, Seoul 08826, Korea\label{i:snu1}
       \and Astronomy program, Dept. of Physics \& Astronomy, SNU, 1 Gwanak-rho, Gwanak-gu, Seoul 08826, Korea\label{i:snu2}
       \and V. N. Karazin Kharkiv National University, 4 Svobody Sq., Kharkiv, 61022, Ukraine\label{i:kha}
       }
  \date{\dots / \dots}


  \abstract
  {The populations of small bodies of the Solar System (asteroids, comets, Kuiper Belt objects) are used to 
   constrain the origin and evolution of the Solar System. Both their orbital distribution and composition distribution
   are required to track the dynamical pathway from their regions of formation to their current locations.}
   {We aim at increasing the sample of Solar System objects (SSOs) that have multi-filter photometry and compositional taxonomy.}
   {We search for moving objects in the \sm Southern Survey.
   We use the predicted SSO positions to extract photometry and astrometry from the \sm frames.
   We then apply a suite of filters to clean the catalog for false-positive detections.
   We finally use the near-simultaneous photometry to assign a taxonomic class to objects.}
   {We release a catalog of
   \numb{\skmtot} individual observations, consisting of 
   \numb{\skmuniq} known and unique SSOs.
   The catalog completeness is estimated to 
   be about \numb{97}\% down to V=18~mag
   and the purity to be above
   \numb{95}\%
   for known SSOs.
   The near-simultaneous photometry provides either three, two, or a single color that we use to
   classify \numb{\taxgr} SSOs with a scheme consistent with the widely used Bus-DeMeo taxonomy.
   }
   {The present catalog contributes significantly to the sample of asteroids with known surface properties
   (about 40\% of main-belt asteroids down to an absolute magnitude of 16). We will release more observations
   of SSOs with future \sm data releases.}
   
   \keywords{Catalogs; Minor planets, asteroids: general}
   \maketitle


\section{Introduction}

  \indent The small bodies of our Solar System
  (asteroids, comets, Kuiper-belt objects) are the remnants
  of the building blocks that accreted to 
  form the planets.
  Their orbital and compositional distributions hold the record
  of the events that shaped our planetary system
  \citep{2009-Nature-460-Levison, 2014Natur.505..629D, 2015-AsteroidsIV-Morbidelli, 2020MNRAS.492L..56C}.
  
  \indent While the number of known Solar System Objects
  (hereafter SSOs) has increased to over a million, the fraction of
  SSOs with known composition remains limited. 
  Spectroscopy in the visible and near-infrared have been used for decades
  to assert the composition, but the sample remains small
  \citep[several thousands, e.g.,][]{2002Icar..158..146B, 
    2009Icar..202..160D, 2014Icar..233..163F, 2019AJ....158..196D, 2019Icar..324...41B}.
  On the other hand, multi-filter photometry can be used to
  classify SSOs in broad compositional groups, providing less details but on
  large samples
  \citep[up to several hundreds of thousands, e.g.,][]{2001-AJ-122-Ivezic, 2016AA...591A.115P, 
    SergeyevCarry2021}
  
\begin{figure}[t]
  \centering
  \input{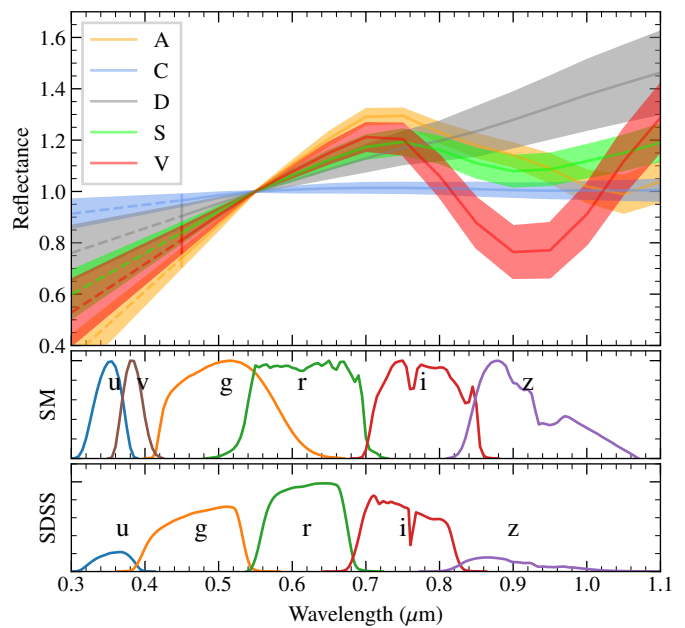}
  \caption{Reflectance spectra of A, C, D, S, and V asteroid classes from 
  \citet{2009Icar..202..160D} normalized at 550nm. \sm filters (transmission curves for a normalized
  quantum efficiency of the CCD are shown in SM panel) are
  well-adapted to spectral characterization. We also report SDSS filters for
  comparison.}
  \label{fig:filter}
\end{figure}

  \indent We analyze here the images from the \sm Southern Survey (Minor Planet Centre (MPC) observatory code \textit{Q55}), 
  which uses a suite of filters well-adapted to asteroid spectral characterization
  (\Autoref{fig:filter}),  similar to those
  of the Sloan Digital Sky Survey
  \citep[SDSS, the main source of compositional information over almost two decades;][]{2004-MNRAS-348-Szabo, 2005-Icarus-173-Nesvorny, 2010AA...510A..43C, 2013Icar..226..723D, 
     2014-Icarus-229-DeMeo,2019Icar..322...13D,
     2008-Icarus-198-Parker, 
  2018-Icarus-304-Graves}. 
  While the SDSS finished its imaging survey in 2009, \sm has started
  its operations in 2014 and is 
  currently active.
  
  \indent The present article aims to increase the number of asteroids
  with multi-filter photometry and taxonomy by identifying
  known SSOs in the \sm source catalog. 
  The article is organized as follows.
  In \Autoref{sec:sm}, we summarize the characteristics of the \sm Southern Survey.
  In \Autoref{sec:extract}, we describe how we
  extract SSOs observations from the \sm point-source
  catalog, and detail the filters applied to the sample
  to reject false-positive sources in \Autoref{sec:clean}.
  The completeness and purity of the catalogue in
  estimated in \Autoref{sec:purity}.
  We present the catalog of SSO colors in 
  \Autoref{sec:catalog}, and use it to
  classify the SSOs consistently with the 
  \citet{2009Icar..202..160D}
  taxonomic classification in \Autoref{sec:taxo}.
  We present our plan for future releases in \Autoref{sec:future}.
  Finally, we summarize the released asteroid sample in
  \Autoref{sec:conclusion}.

\section{The \sm survey\label{sec:sm}}

  The \sm Southern Survey (SMSS) is producing a homogeneous multi-band atlas
  of the whole Southern Hemisphere in $u,v,g,r,i,z$ filters
  \citep{2018PASA...35...10W}. Observations were carried out with a 1.35m telescope located at Siding Spring Observatory (IAU code Q55). The telescope has an f/4.8 focal ratio and  is equipped with a mosaic CCD camera having 268 million pixels.
  The third data release (DR3) covers an area of more than 24,000 deg$^2$ and contains
  over 200,000 images with over 8 billion individual source detections
  \cite[see the SkyMapper website\footnote{See \url{https://skymapper.anu.edu.au}} and the
  DR2 release article for details:][]{2019PASA...36...33O}.
  The data in DR3 were obtained between March 2014 and October 2019.
  Individual deep exposures can reach magnitudes of 20 in $u,v,z$, 21 in $i$,
  and 22 in $g,r$ (10$\sigma$ detections) in the AB 
  system \citep{1983ApJ...266..713O}. The median seeing ranges from 3.3\arcsec in $u$ to 2.5\arcsec in $z$.
  
  The SMSS covers each field on the sky in three primary modes:
  a shallow 6-filter sequence with exposure times between 5 and 40 seconds that reaches depths of 18~ABmag,
  a deep 10-image sequence of $uvgruvizuv$ with 100-second exposures,
  and pairs of deep exposures in $gr$ and $iz$.
  This observing strategy, in conjunction with the enhanced sensitivity of $g,r$,
  gives rise to a predominance of $g-r$ colors in the results presented in \Autoref{sec:catalog},
  but almost always leads to the measurement of at least one photometric color obtained with
  $\lesssim 2$ minutes between exposure midpoints in the two filters.
 
\begin{figure*}[h]
    \centering
    \includegraphics[width=1.0\hsize]{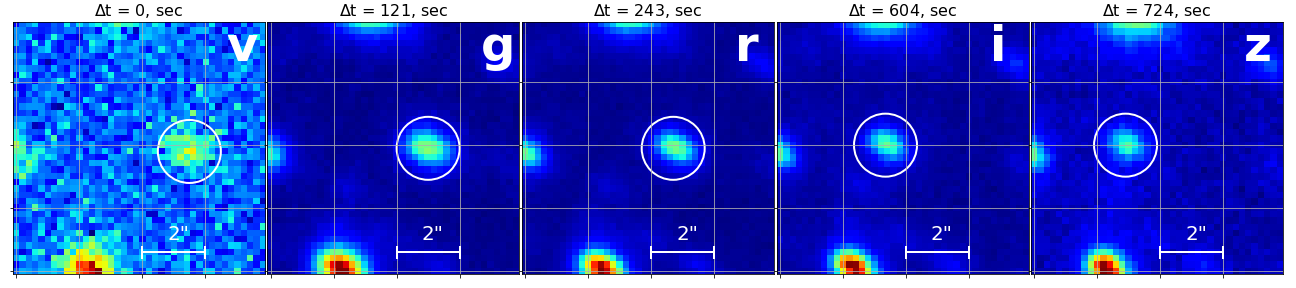}
    \caption{\sm multi-color observations of SSOs, here the asteroid (4365) Ivanova.
    The time interval between each frame and the first one (here $v$) is reported above each image.
    North is up and East is left.}
    \label{fig:ast_example}
  \end{figure*}

  The work presented here adopts a single photometric measurement associated with
  each \sm image, although for objects with significant motion during the exposure,
  additional information on shape and rotation parameters may be available
  from a more detailed analysis \cite[e.g., the \sm observations of the Earth-impacting 2018~LA,][]{2021MPS...56..844J}.
  We present a typical suite of images illustrating the apparent motion of SSOs in \sm frames
  (\Autoref{fig:ast_example}).

\section{Extracting candidate SSOs\label{sec:extract}}

  For each of the 208,860 images contained in SMSS DR3, 
  we compile
  all known SSOs potentially present in the images by performing a search with \skybot
  \citep{2006-ASPC-351-Berthier, 2016-MNRAS-458-Berthier}, 
  a Virtual Observatory Web Service
  providing a cone-search utility for Solar System objects. 
  The 2\degr\ cone-search radius utilised is slightly larger than the 1.7\degr\
  centre-to-corner size of the \sm camera and returned \numb{\initsrc} predicted SSOs locations
  with no initial filtering with a magnitude limit.
  Among these, \numb{\onccd} are predicted within the field of view. The smaller number in the number of SSO predicted locations is explained by the search and FoV area difference.
  
  For each predicted SSO position, 
  we extract all sources listed in the SMSS DR3 \texttt{photometry}
  table\footnote{The SMSS \texttt{photometry} table contains the per-image measurements, as opposed to the averaged 
  quantities in the \texttt{master} table.}
  within a radius
  of twice the \skybot-reported position uncertainty (with a floor of \numb{5}\arcsec).
  We obtain a list of 
  \numb{\skmsrc} \sm sources, associated with the
  \numb{\skmmatch} predicted SSO positions.

\begin{figure}[t]
  \centering
  \includegraphics[width=.95\hsize]{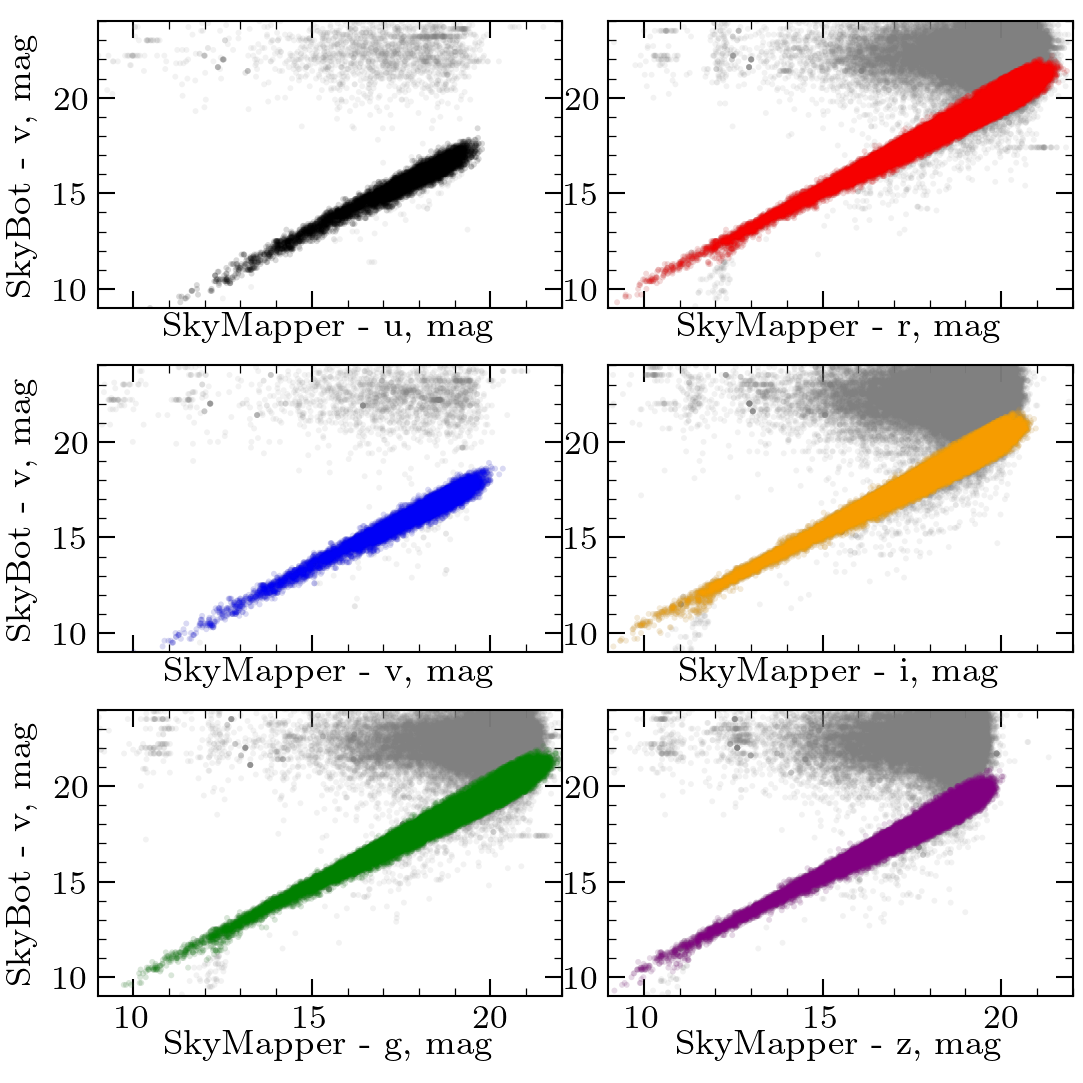}
  \caption{Measured \sm magnitude in each filter compared
   with the predicted V magnitude of SSOs (colored dots).
   The light grey dots correspond to rejected sources, spuriously associated
   to SSOs too faint to have been detected (see text).}
  \label{fig:dmag}
\end{figure}

\begin{figure}[t]
  \centering
  \includegraphics[width=.95\hsize]{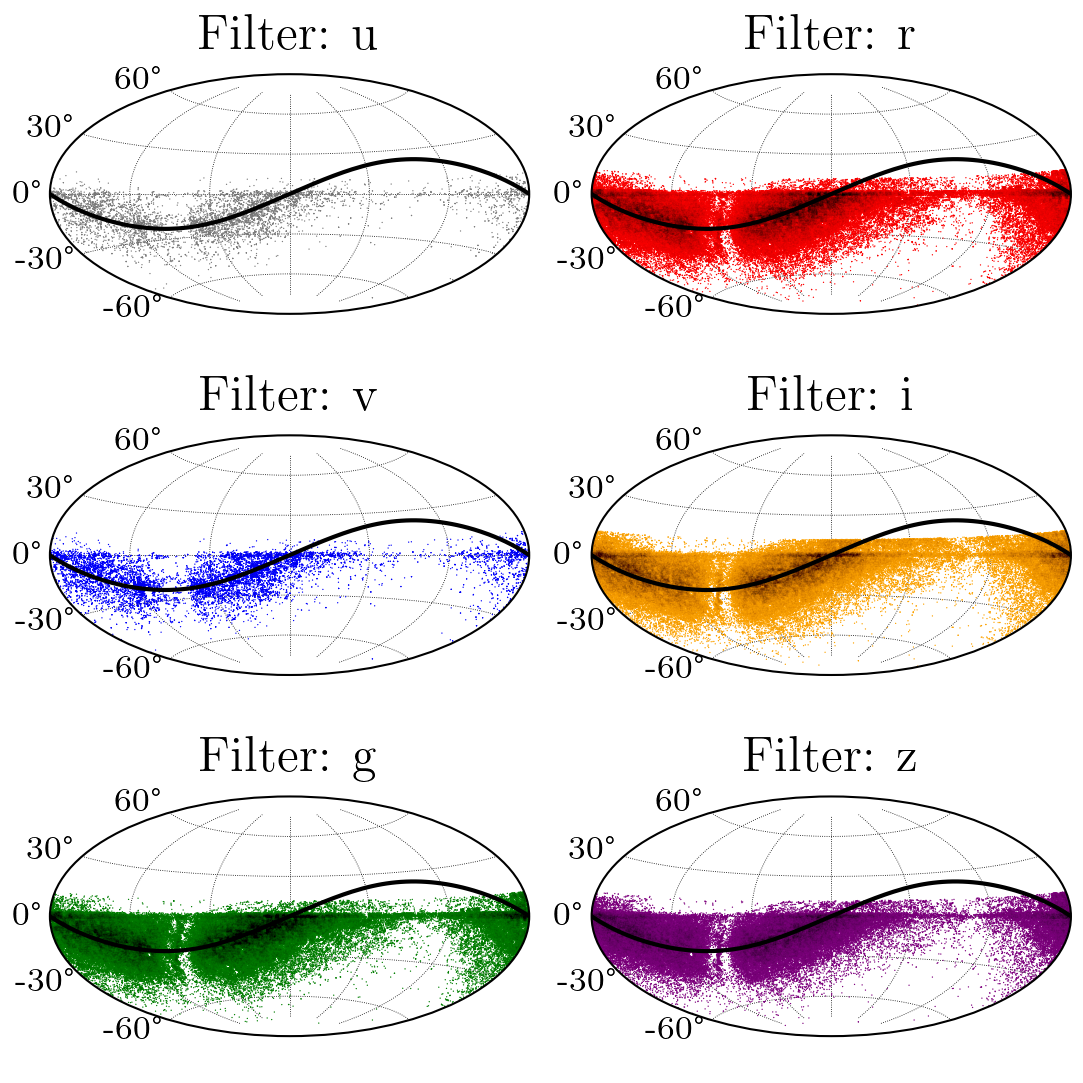}
  \caption{Sky distribution (equatorial frame) of 880,528 SkyMapper SSO observations in each filter.
    The black curve represents the ecliptic plane.}
  \label{fig:sky}
\end{figure}

\section{Rejecting false-positive sources\label{sec:clean}}

  The fraction of interlopers among the \numb{\skmsrc} sources extracted above 
  is large. First, multiple \sm sources may be located by mere chance 
  in close vicinity to predicted SSO positions. Second, many SSOs are too faint to have been 
  imaged by \sm, and the extracted sources correspond to spurious associations.
  We thus apply \numb{two} filters to the extracted sources to reject false-positive SSOs associations.
  \subsection{Comparison with Gaia}

    We first identify stationary sources (stars, unresolved galaxies) by
    comparing the catalog of sources with Gaia DR2
    \citep{2018AA...616A...1G}.
    We find \numb{\skmgaia} Gaia sources within \numb{2}\arcsec
    (less than the median \sm seeing, \Autoref{sec:sm})
    of the extracted
    sources.
    Although this certainly includes some real associations with observed SSOs, we remove them from the catalog
    to avoid introducing biases in the measured photometry.

  \subsection{Comparison with the expected photometry}

    We present in \Autoref{fig:dmag} a 
    comparison of the predicted V magnitudes of SSOs with \sm-measured magnitudes. 
    These magnitudes are in almost all cases the PSF (Point Spread Function) photometry.
    However, owing to the apparent motion of SSOs during an exposure, the PSF photometry may underestimate the true magnitude of the observed SSO. 
    We computed this underestimation as a function of exposure time and SSO apparent velocity.
    We report PSF magnitudes for most SSO observations, only reporting Petrosian magnitudes
    for observed SSOs with an apparent velocity
    of more than 50 \arcsec/hour and image exposure time of 100\,s. PSF photometry was preferred
    over Petrosian overall due to its higher accuracy in the case of non-trailed sources.
    
    Many contaminants are still present, and easily identified by their large 
    magnitude difference with respect to the predicted magnitudes.
    These contaminants correspond to SSOs too faint to have been detected, and
    erroneously associated to 
    stationary sources.

    As these contaminants tend to be located at large angular separation from the predicted
    SSO position, we reject all
    sources farther than \numb{5}\arcsec from \skybot prediction 
    (which also implies a high \skybot position uncertainty). 
    The next cleaning procedure was based on the \skybot and \sm magnitude difference. We compared visual magnitudes as predicted by \skybot with the \sm sample and calculated regression slopes and offsets for each filter, which are presented in the \Autoref{tab:regression}. Then we removed sources showing a difference in magnitude minus the photometry magnitude error 
    larger than \numb{1.0} mag.
    A few observed SSOs appear to have saturated ($g$,$r$,$i$ filters),
    with magnitudes $\lesssim 10$ in shallow images or $\lesssim 13.8,13.8,13$ in deeper images,
    and the latter condition 
    removes them too.
    
    \begin{table}[t]
        \centering
        \begin{tabular}{crrr}
    \hline\hline
    Filter & Slope & Offset & \#$_{\textrm{SSOs}}$ \\
    \hline
    u & 0.927 & -0.776 & 2850 \\
    v & 0.937 & -0.550 & 5302 \\
    g & 0.976 & 0.133 & 129789 \\
    r & 0.973 & 0.571 & 151809 \\
    i & 0.971 & 0.814 & 125395 \\
    z & 0.968 & 0.837 & 62774 \\
    \hline
\end{tabular}
        \caption{Average values of the slope
        and the magnitude offset between \skybot SSOs visual magnitude
        and each \sm band magnitude.}
        \label{tab:regression}
    \end{table}
    
    Finally, whenever several \sm sources remain associated with a SSO, we select the closest and reject the others.
    After these steps of filtering, we 
    obtained a list of \numb{\skmtot} measurements of \numb{\skmuniq} individual SSOs.
    We present their distribution on sky in \Autoref{fig:sky}.
    
\begin{figure}[t]
    \centering
    \input{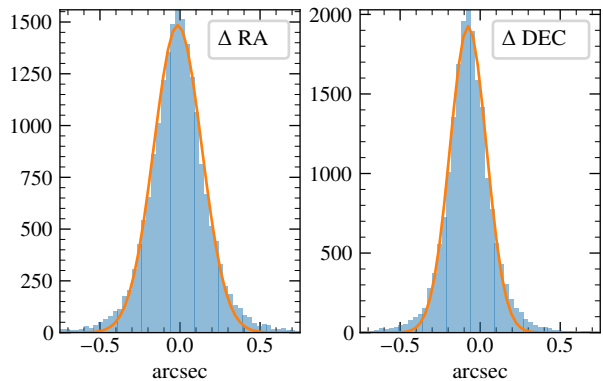}
    \caption{The distribution of the coordinate difference between \sm sources and predicted by \skybot SSO positions.}
    \label{fig:centerdist}
\end{figure}
\begin{figure}
      \centering
      \includegraphics[width=.95\hsize]{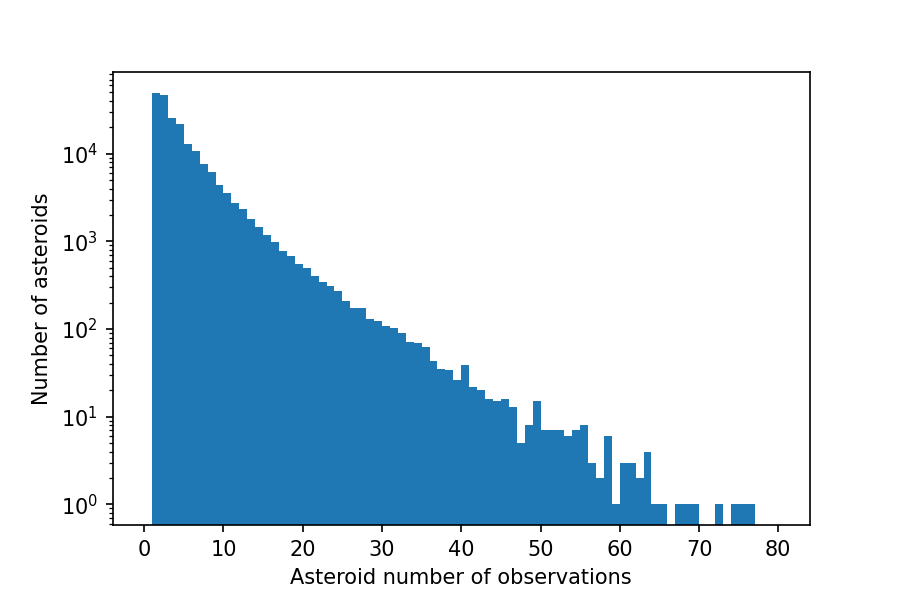}
      \caption{The \sm distribution of asteroid observations.}
      \label{fig:observ_dist}
  \end{figure}

    The mean difference between predicted and measured SSO positions are -0.014 and -0.075 arcsec with standard deviations of 0.15 and 0.11 arcsec for RA and DEC respectively (\Autoref{fig:centerdist}), well below the typical seeing of \sm images. 
    Though the mean number of observations for an individual SSO is about four,
    hundreds of asteroids have dozens of observations, while most of them were observed only one or two times see \autoref{fig:observ_dist}. 
    We present the distribution of measurements among dynamical classes in \Autoref{tab:dyncls}, and describe the catalog of measurements in \Autoref{app:cat}.

\begin{table}[t]
    \centering
    \caption{Number of 
    observations (N$_{\textrm{obs}}$) of 
    N$_{\textrm{obj}}$ unique objects, sorted by dynamical classes.
    NEA, MBA, and KBO stand for
    Near-Earth Asteroid, 
    Main-belt Asteroid, 
    and 
    Kuiper Belt Object.}
    \label{tab:dyncls}
    \begin{tabular}{lrr}
    \hline \hline
    Dynamical class & N$_{\textrm{obj}}$ & N$_{\textrm{obs}}$ \\
    \hline
    NEA>Aten & 39 & 103 \\ 
    NEA>Apollo & 282 & 822 \\ 
    NEA>Amor & 348 & 1289 \\ 
    Mars-Crosser & 2,487 & 9,807 \\ 
    \hline        
    Hungaria & 4,035 & 13,803 \\ 
    MBA>Inner & 65,231 & 287,664 \\ 
    MBA>Middle & 71,677 & 301,935 \\ 
    MBA>Outer & 57,721 & 245,664 \\ 
    MBA>Cybele & 811 & 4,481 \\ 
    MBA>Hilda & 829 & 4,298 \\ 
    \hline        
    Trojan & 1,928 & 10,103 \\ 
    Centaur & 24 & 127 \\ 
    KBO & 37 & 189 \\ 
    Comet & 65 & 226 \\ 
    Planet & 1 & 17 \\ 
    \hline        
    Total & 205,515 & 880,528 \\ 
    \hline
    \hline
\end{tabular}
\end{table}

\section{Purity and completeness\label{sec:purity}}

  \indent The completeness indicates the fraction of reported SSO observations with respect to how many SSOs were present in the field of view.
  The purity indicates the fraction of contamination among the released observations. 
  For both estimators, the closer to unity the better.
  We estimate the completeness by comparing the number of SSO observations predicted by \skybot
  with the number of 
  sources after filtering. 

\begin{figure}[t]
    \centering
    \input{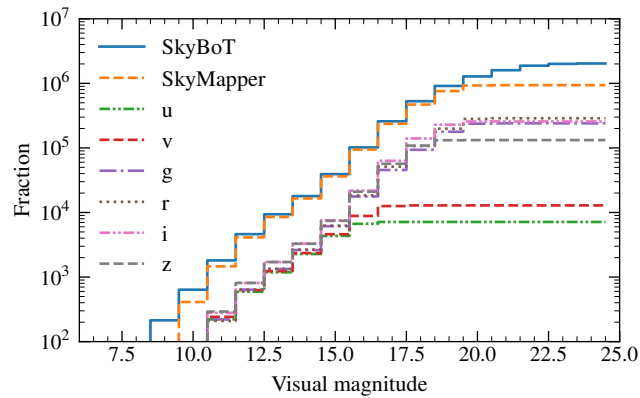}
    \caption{Completeness as function of predicted V magnitude for the whole
    catalog (\sm), and for each filter. 
    }
    \label{fig:completeness}
\end{figure}

    At face value (\numb{\onccd} vs \numb{\skmtot}), the completeness is \numb{5\%} only. Most non-detected SSOs are, however, simply those beyond 
    the \sm field of view, or too faint to be detected.
  Thus the completeness of the sources inside the CCD field of view for V between 11 and 18 mag is above \numb{97}\%.
  (\Autoref{fig:completeness})
  
  \indent We estimated the purity through a visual inspection of frames centered
  on SSO positions in both \sm and Pan-STARRS (Panoramic Survey Telescope And Rapid Response System)
  archives.
  We generated cut-out frames using the \sm image
  cutout service\footnote{See \url{https://skymapper.anu.edu.au/how-to-access/}}, and
  downloaded Pan-STARRS cut-out stacked images at the same coordinates
  \citep{2016arXiv161205560C}.
  The Pan-STARRS survey has a significantly deeper limiting magnitude. Thus, all stationary \sm
  sources should hence be visible in Pan-STARRS images.
  As the Pan-STARRS survey did not image the sky below
  -30\degr of declination, we test the purity for sources located North
 of this declination only.
  
 \begin{table}[t]
    \centering
    \caption{Purity (expressed in percent) of MBAs, NEAs, and KBOs, as function
    of their apparent V magnitude. We also the number of sources ($N$) we visually check
    in each magnitude bin.}
    \label{tab:purity}
    \begin{tabular}{ccccccc}
\hline \hline
V & $N_{MBA}$ & $N_{NEA}$ & $N_{KBO}$ & MBA & NEA & KBO \\
(mag) &&&& (\%)& (\%)& (\%) \\
\hline
10 &0 & 0 & 0 &- &- &- \\
11 &2 & 5 & 0 &100 &100 &- \\
12 &2 & 0 & 0 &100 &- &- \\
13 &4 &23 & 0 &100 &100 &- \\
14 &6 & 5 & 0 &100 &100 &- \\
15 & 14 &23 &49 &100 &100 &100 \\
16 & 40 &83 & 0 &100 & 99 &- \\
17 &105 & 247 & 0 &100 &100 &- \\
18 &174 & 321 &14 &100 &100 &100 \\
19 &289 & 421 &29 &100 &100 &100 \\
20 &298 & 418 &54 & 99 & 99 &100 \\
21 & 63 & 133 &30 & 94 & 93 & 97 \\
22 &2 & 4 & 4 & 50 &100 & 75 \\
23 &0 & 0 & 0 &- &- &- \\
\hline
\end{tabular}

\end{table}
  
  We inspected all available \numb{189} KBOs (Kuiper Belt Objects) observations,
  \numb{1,683} NEAs (Near Earth Asteroids), and 
  \numb{1,000} randomly selected MBAs (Main Belt Asteroids) observations with declination of greater than -30 deg.
  The purity of the \sm SSOs survey is about \numb{100}\% down to magnitude
  \numb{V$\approx$20}, where it begins to drop (\Autoref{tab:purity}).

\section{Computation of colors\label{sec:catalog}}

\begin{figure}[t]
  \centering
  \input{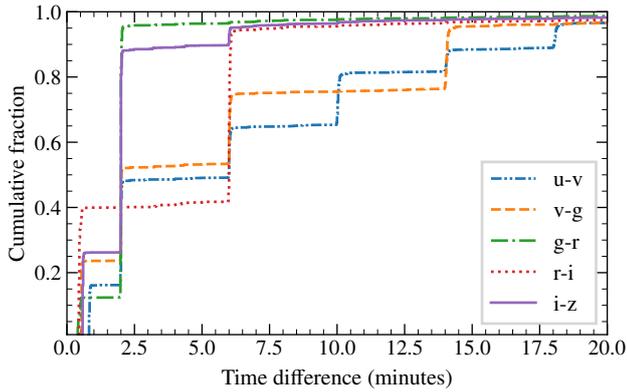}
  \caption{Cumulative histogram of the time difference between filters.}
  \label{fig:cumm_hist}
\end{figure}

We build the colors of the detected SSOs for compositional (taxonomic) purposes.
Due to the intrinsic photometric variability of asteroids caused by their irregular shape \citep{2004-MNRAS-348-Szabo, 2016AA...591A.115P, 2018AA...609A.113C},
color indices have to be calculated from near-simultaneous observations.
We only consider observations taken within a maximum of 20 minutes of each other.
Although this threshold may appear large, it mainly affects $u$ and $v$ filter combinations.
The majority (\numb{95+\%}) of $g$-$r$ and $i$-$z$ colors were 
acquired within \numb{2} minutes, and $r$-$i$ in \numb{6} minutes
(\Autoref{fig:cumm_hist}). 

\citet{2004-MNRAS-348-Szabo} analyzed the color variability of asteroids from the 4th SDSS Moving Object Catalog (MOC4), and with conservative assumptions, showed that 5-minute time differences have an effect on the color of less than 0.03 mag. Therefore, the typical 2-minute difference in our $g-r$ and $i-z$ data will not have a significant effect on the color estimation.

We provide the catalog of \numb{\colorsmes} measured colors of \numb{\indssowithcolors} SSOs, specifying the time difference between each acquisition, and describe the catalog elements in \Autoref{app:cat}.

Some asteroids had their colors measured multiple times.
  We thus calculated the weighted mean colors value of individual SSOs, 
  taking into account the magnitude uncertainty and the time difference between
  observations and set weights as: $1/mag_{err} + 0.1/\Delta d$,
  where $\Delta d$ is the time difference in days while $mag_{err}=\sqrt{mag1_{err}^2 + mag2_{err}^2}$ is color uncertainty of two photometry measurements $mag1, mag2$. If the SSO has multiple color measurements the error was computed as the weighted mean value of color uncertainties.
  We also computed \numb{\colorcalc} colors that were not directly observed, but are combinations of measured colors: for instance, we compute $r-z$ from $g-r$ and $g-z$.
  
  As a result, we constructed a catalog of \numb{\indssowithcolors} SSOs which contain at least one measured color. The total number of unique SSO colors is \numb{\colorsindsso}, both observed and derived
  (\Autoref{tab:color_stat}).

\begin{table}[t]
    \centering
    \caption{Number of measured colors (N$_{\textrm{mes}}$) for each pair of filters, 
    associated with N$_{\textrm{obj}}$ unique SSOs. We also report the number of 
    colors computed from linear combination of colors (N$_{\textrm{comp}}$, see text).}
    \label{tab:color_stat}
    \begin{tabular}{crrr}
    \hline \hline
    Color & N$_{\textrm{mes}}$ & N$_{\textrm{obj}}$ & N$_{\textrm{comp}}$ \\
    \hline
    g-r   &  201,910 &  117,356 &     205 \\
    i-z   &  130,065 &   57,735 &     186 \\
    r-i   &   86,518 &   58,626 &   1,425 \\
    g-i   &   74,693 &   54,759 &   2,737 \\
    r-z   &   55,560 &   36,689 &   3,383 \\
    g-z   &   48,290 &   34,333 &   4,066 \\
    u-v   &   12,028 &    2,628 &      26 \\
    v-g   &    9,963 &    4,316 &     211 \\
    v-r   &    9,813 &    4,317 &     245 \\
    v-z   &    9,461 &    4,241 &     306 \\
    v-i   &    9,396 &    4,254 &     331 \\
    u-g   &    5,620 &    2,432 &     139 \\
    u-r   &    5,551 &    2,434 &     148 \\
    u-z   &    5,363 &    2,410 &     176 \\
    u-i   &    5,314 &    2,413 &     199 \\
    \hline
    Total &  669,545 &  388,943 &  13,783 \\
    \hline
\end{tabular}
\end{table}

  In order to check the time lag effect on the \sm asteroids colors, we estimated the dependence of brightness variability on asteroids rotation period. We downloaded the latest data release from the \href{https://minplanobs.org/MPInfo/php/lcdb.php}{Asteroid Lightcurve Database}\footnote{See \url{https://minplanobs.org/MPInfo/php/lcdb.php}} which contains periods and amplitudes of more than 30,000 known SSOs. Then we compared \sm data and selected joint asteroids. 
  We obtained samples of more than twelve thousands asteroids with $g-r$ and $i-z$ colors and almost eight thousand asteroids with $g-i$ colors.
  For each asteroid in the samples, we calculated the 
  maximum expected color change arising from asteroid rotation between the \sm imaging epochs (still limited to a 20-minute window) as the product of the photometric amplitude and the number of half-periods represented by the \sm time difference.


  The magnitude difference in our samples from the rotation effect was less than the color uncertainties for 96, 97, and 87 percent of asteroids in $g-r$, $i-z$, and $g-i$ colors, and thus we consider the overall impact of rotation on the colors we report to be small.

\section{Taxonomy\label{sec:taxo}}

  We used the multi-color photometry to classify asteroids in a scheme consistent
  with the widely used Bus-DeMeo taxonomy \citep{2009Icar..202..160D} in which asteroid color values are used to define belonging to a certain taxonomy complex.
  Following our recent work with the SDSS \citep{SergeyevCarry2021}, 
  we use a modified version of the approach of \citet{2013Icar..226..723D} which a decision tree based on colors is used to assign the taxonomic class.
  Where, for the each asteroid, we calculate the probability of it being associated with each taxonomic broad complex (A, B, C, D, K, L, Q, S, V, and X). We computed the intersection between the volume occupied by each taxonomy complex and the color(s) of the asteroid, represented as a $n$-dimensional Gaussian probability density function based on color values and its uncertainties. \citep{SergeyevCarry2021}.

  \subsection{\sm taxonomy boundaries}

\begin{figure}[t]
    \centering
    \input{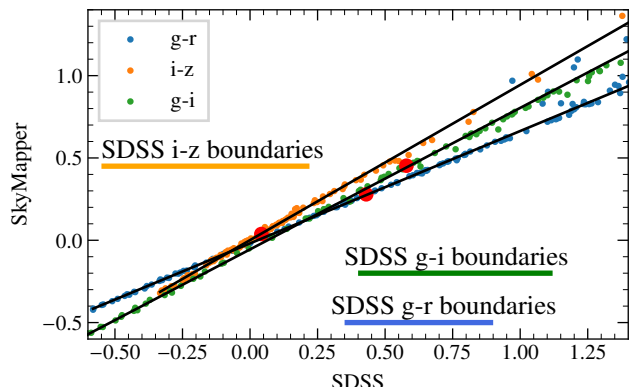}
    \caption{The $g-r$ (blue), $g-i$ (green), and $i-z$ (orange) colors
    from the SDSS compared with \sm for a wide range of stellar
    classes
    Red dots represent the colors of the Sun \citep{2006MNRAS.367..449H}.
    The colored horizontal lines illustrate the range of colors for the
    SDSS asteroids taxonomy \citep{SergeyevCarry2021}.
    }
    \label{fig:transform_color}
\end{figure}

    We first convert the color ranges
    of each taxonomic class \citep{2013Icar..226..723D}
    from SDSS to \sm filters.
    We used the color coefficients of \citet{2019MNRAS.482.2770C} and \citet{2019ApJS..243....7H}.
    We compared the $g-r$, $g-i$, and $i-z$ values from \sm with those from the SDSS for
    a wide range of stellar classes
    (\Autoref{fig:transform_color}, data from \citet{1998PASP..110..863P}).
    We fitted the color-color dependency by linear regression using the
    \citet{10.1093/biomet/69.1.242} approximation.
    These linear coefficients were used to convert the SDSS color boundaries into the \sm 
    photometric system.

\subsection{Multi-color based taxonomy}

\begin{figure*}[t]
    \centering
    \includegraphics[width=0.95\hsize]{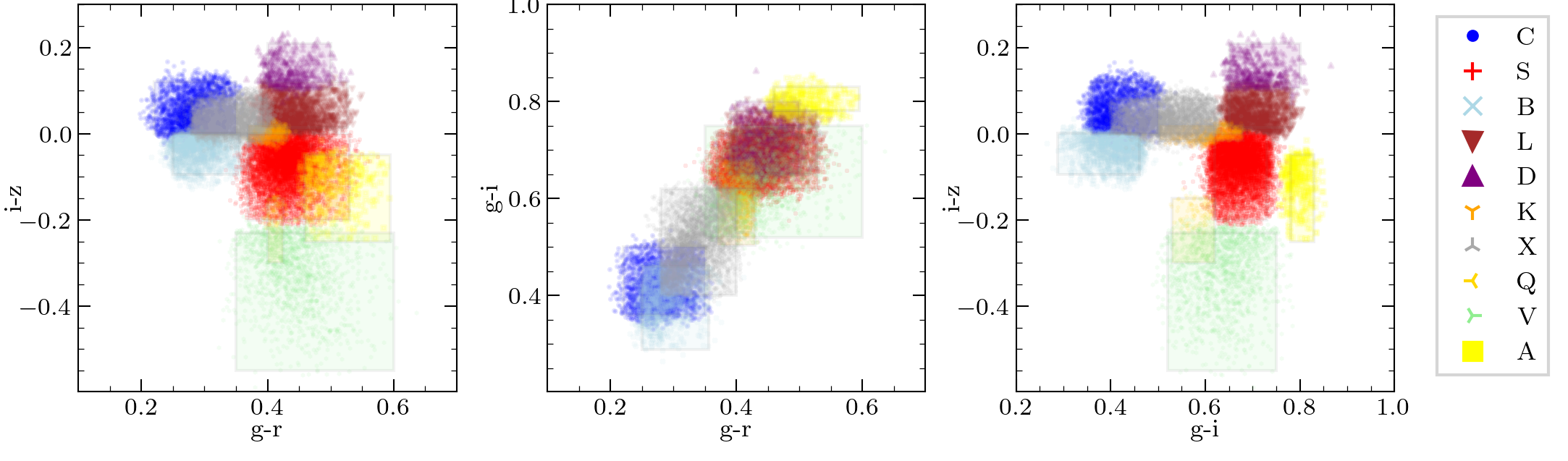}
    \caption{Taxonomy of \numb{29,779} \sm asteroids with a probability larger than \numb{0.2} with three colors.
    Boxes show the boundaries of the taxonomic classes. Color points mark individual asteroids.}
    \label{fig:tax_color}
 \end{figure*}
 
 \begin{figure*}[t]
    \centering
    \input{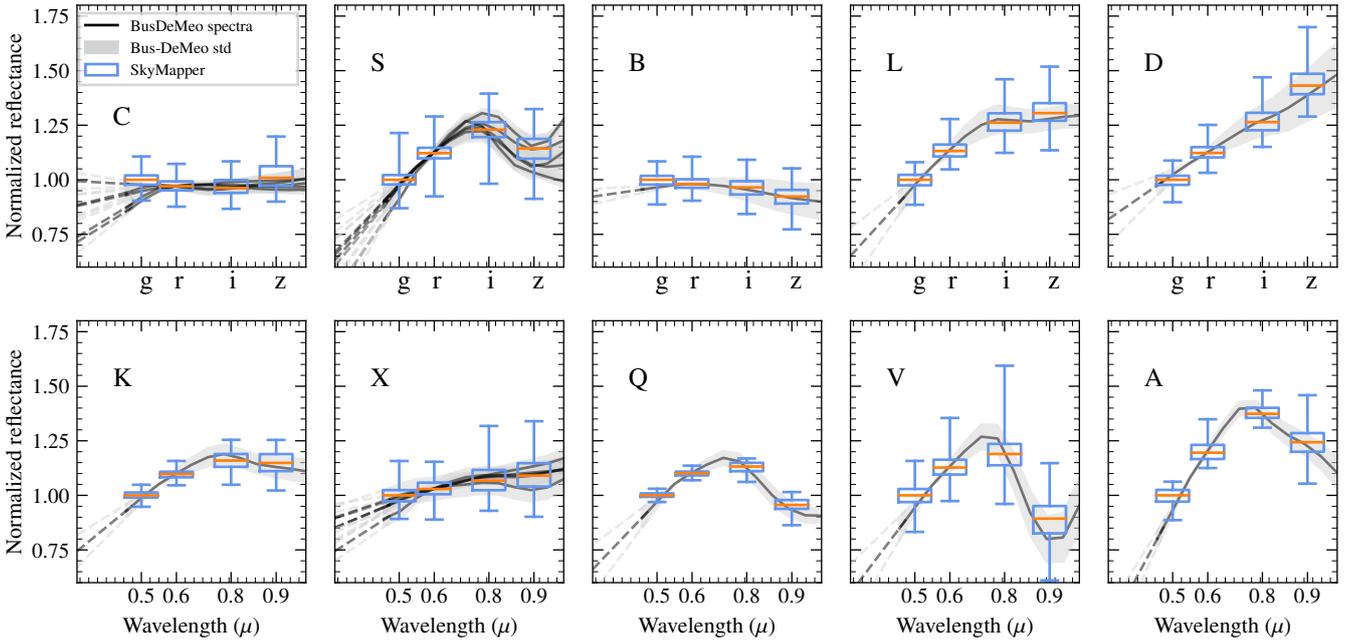}
    \caption{Pseudo-reflectance spectra of asteroids 
      based on their $g-r$, $g-i$, and $i-z$ colors. We indicate the average wavelength of each filter in the lower plots.
      The distribution of values for each band is
      represented by whiskers (95\% extrema, and the 25, 50, and 75\% quartiles).
      For each class, we also represent
        the associated template spectra of the Bus-DeMeo taxonomy
        \citep{2009Icar..202..160D}.}
    \label{fig:grgiiz_reflectance}
\end{figure*}

 Owing to the survey strategy, not all SSOs have the same suite of colors
  (\Autoref{tab:color_stat}). We thus adapt our approach \citep{SergeyevCarry2021} to handle both
  three-color ($g-r$, $g-i$, $i-z$) and
  two-color -- ($g-r$,$i-z$), ($g-i$,$i-z$), or ($r-i$,$i-z$) -- cases.
  We restrict the list of pairs of colors to those containing $i-z$:
  it probes the 1\,$\mu$m absorption band which is among the most characteristic
  spectral feature asteroid taxonomies \citep{1975-Icarus-25-Chapman}.
  We also shrink the boundaries to more stringent ranges for the cases with two colors only.

  For each observation, we compute the volume it occupies in the 
  color space (either 2D or 3D) based
  on the corresponding Gaussian distribution, whose $\sigma$ are set to color uncertainties.
  We then compute a score for each class, $\mathcal{P}_k$, based on 
  the volume of the intersection between the volume of each observation 
  and the space occupied by each taxonomic complex (\Autoref{fig:tax_color}), 
  normalized by the volume of the Gaussian:
  
  \begin{equation}
    V_{\sigma} = \prod_{j=1}^N{
    \left(erf\left[\frac{b_j-\mu_j}{\sqrt{2}\sigma_j}\right] - erf\left[\frac{a_j-\mu_j}{\sqrt{2}\sigma_j}\right] \right)}
  ,\end{equation}
  
  \noindent where
  $erf(z)$ is the error function,  $erf(z) = \frac{2}{\sqrt{\pi}}\int_{0}^{z}{e^{-t^2}dt }$, 
  the index $j$ indicates the 
  colors (with $N \in \{2,3\}$),  
  $a_j$ and $b_j$ are the color boundaries of the complexes, and
  $\mu_j$ and $\sigma_j$ are the color and uncertainty of the SSO. Hence, 
  for a given observation, the volumes of all intersections sum to one.
  These normalized volumes correspond to the probabilities
  $\mathcal{P}_k$ of pertaining
  to each taxonomic class.

\begin{figure*}[]
  \centering
  \includegraphics[width=.90\hsize]{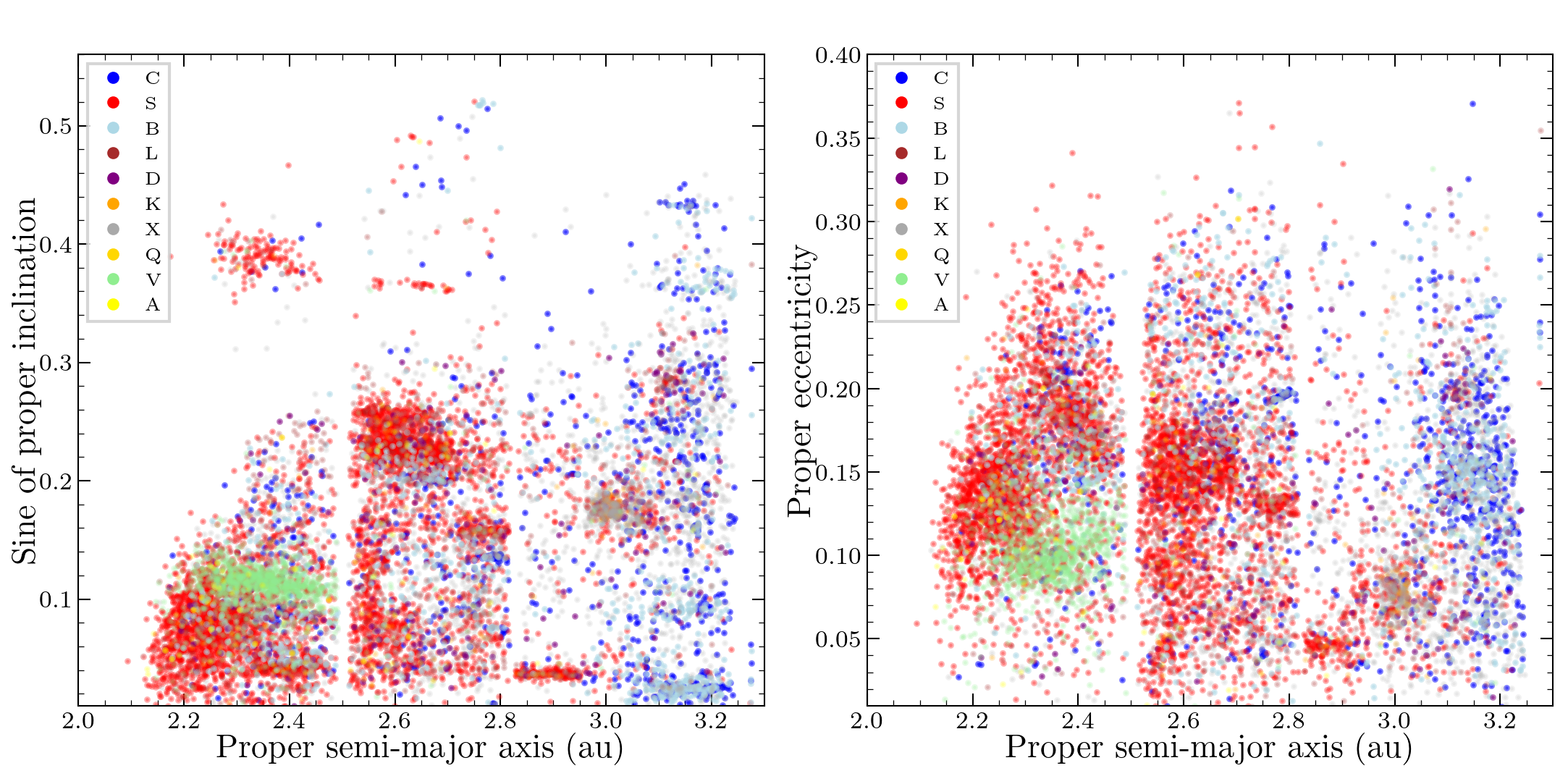}
  \caption{Orbital distribution of the SSOs, color-coded by taxonomic class based on their $g-r$, $g-i$, and $i-z$ colors. 
  }
  \label{fig:taxo_regions_grgiiz}
\end{figure*}

\begin{figure}[]
    \centering
    \input{figs/grgiiz_matrix.pgf}
    \caption{Confusion matrix of the taxonomy of \numb{1,697} obtained by spectra asteroids and from Skymapper observations based on their $g-r$, $g-i$, and $i-z$ colors. The values are reported in percent.}
    \label{fig:matrix_grgiiz}
\end{figure}

  We then assign to each object its most probable class.
  The only exception to that rule is the unknown class (labeled U), which is assigned only 
  if its probability is strictly equal to 1. Otherwise, whenever U is the most probable but not equal to unity, 
  we assigned the second most-probable class.
  As an example, we present in \Autoref{fig:tax_color} the distribution in the three colors space
  of \numb{29,779} asteroids for which the probability is above 0.2
  (out of \numb{\taxgrgiiz} asteroids with three colors).
  Their
  pseudo-reflectrance spectra in \Autoref{fig:grgiiz_reflectance}
  \citep[computed using the solar colors of][adapted to \sm]{2006MNRAS.367..449H}
  show a good agreement with
  the template spectra of taxonomic classes \citep{2009Icar..202..160D}.

  We illustrate the potential of this taxonomic classification in \Autoref{fig:taxo_regions_grgiiz} the photometric classes show clear concentrations in the orbital parameters for different asteroid families. For example asteroids of the Vesta family (light green points) are concentrated in the inner belt which is otherwise dominated by asteroids of S complex (red points).
C and B complexes dominate in the outer belt region, however, the Koronis family at 2.9 AU distance shows a clear S complex taxonomy. 
It is interesting to note that X-type asteroids (indicated by gray points) are a mixture of P, M, and E asteroid types by Tholen classification \citep{1989aste.conf.1139T}, concentrate in the outer belt region associated with the Eos family.

  We test this classification by comparing the \sm classes against previously reported taxonomy
  from spectroscopy \citep[e.g.,][]{2002Icar..158..146B, 2004Icar..172..179L, 2019Icar..324...41B, 2004Icar..172..221F, 2014Icar..233..163F}.
  We found \numb{1,683} common SSOs and compare their classes in \Autoref{fig:matrix_grgiiz}).
  As expected, rare and peculiar classes such as A, K, L and Q may be underestimated here, and misclassified
  as S. Similarly, there is some confusion between C/B and C/X classes, that only differ by spectral slope.
  The values are, however, concentrated on the diagonal of the confusion matrix, showing an overall agreement.
    
  We repeat the same exercise for each SSO with two colors among
  ($g-r$,$i-z$), ($g-i$,$i-z$), or ($r-i$,$i-z$), accounting for
  \numb{\taxgriz}, \numb{\taxgiiz}, and 
  \numb{\taxriiz} SSOs, respectively.
  We present the statistic for all these taxonomic classes in \Autoref{tab:taxonomy}.
  
\begin{table}[]
    \centering
    \caption{Number of asteroids in each taxonomic complex, for different color sets.}
    \label{tab:taxonomy}
    \begin{tabular}{crrrrr}
    \hline \hline
Complex & ($gr,gi,iz$) & ($gr,iz$) & ($gi,iz$) & ($ri,iz$) & ($gr$) \\
\hline 
A     &     537 &     309 &     947 &   4,551 &         \\
B     &   1,914 &   4,587 &   2,760 &   1,317 &         \\
C     &   2,180 &   6,175 &   2,735 &   3,554 &   40,573 \\
D     &     466 &     901 &   1,544 &   1,404 &         \\
K     &   1,005 &   1,254 &   2,661 &     882 &         \\
L     &   1,709 &   3,611 &   1,275 &     973 &         \\
Q     &     164 &     633 &     845 &     532 &         \\
S     &   8,256 &  13,753 &  10,577 &  10,608 &   76,783 \\
V     &   1,787 &   1,125 &     843 &   1,573 &         \\
X     &   4,876 &   2,398 &   4,448 &   5,479 &         \\
U     &   9,882 &   4,317 &   4,664 &   4,625 &         \\
\hline
Total &  32,776 &  39,063 &  33,299 &  35,498 &  117,356 \\
\hline
\end{tabular}

\end{table}

\subsection{Single-color based taxonomy\label{ssec:gr}}

  \begin{figure}[]
    \centering
    \input{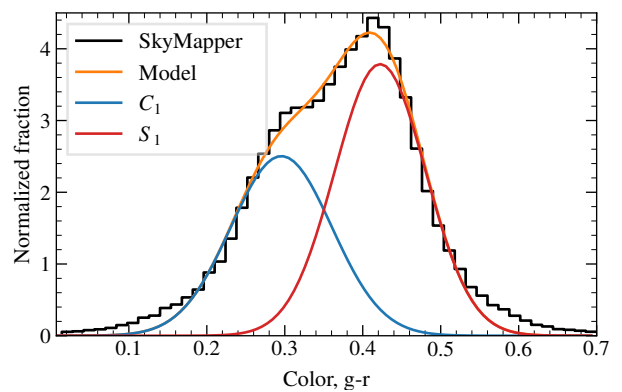}
    \caption{The $g-r$ color distribution of \sm asteroids (black) where fitted by the sum of two normal distributions (orange) which could be associated with $C_1$ (opaque-rich, blue) 
    and $S_1$ (mafic-silicate rich, red) asteroids complexes.
    }
    \label{fig:gr_distribution}
  \end{figure}

    A large number of \numb{\taxgr} SSOs in \sm have only one color: $g-r$.
    A detailed classification cannot be achieved, but this color can 
    still be used to split asteroids into mafic-silicate rich and opaque-rich.
    Similarly to the work by \citep{2020ApJS..247...13E},
    we thus classed the asteroids into
    S-like and C-like objects (hereafter labeled $S_1$ and $C_1$).
    The $g-r$ colors of asteroids clearly present a bi-modal distribution, 
    well-represented by two normal distributions
    (\Autoref{fig:gr_distribution}) corresponding to the
    $C_1$ (bluer) and $S_1$ (redder) classes.
  
  \begin{figure*}[]
  \centering
  \includegraphics[width=.95\hsize]{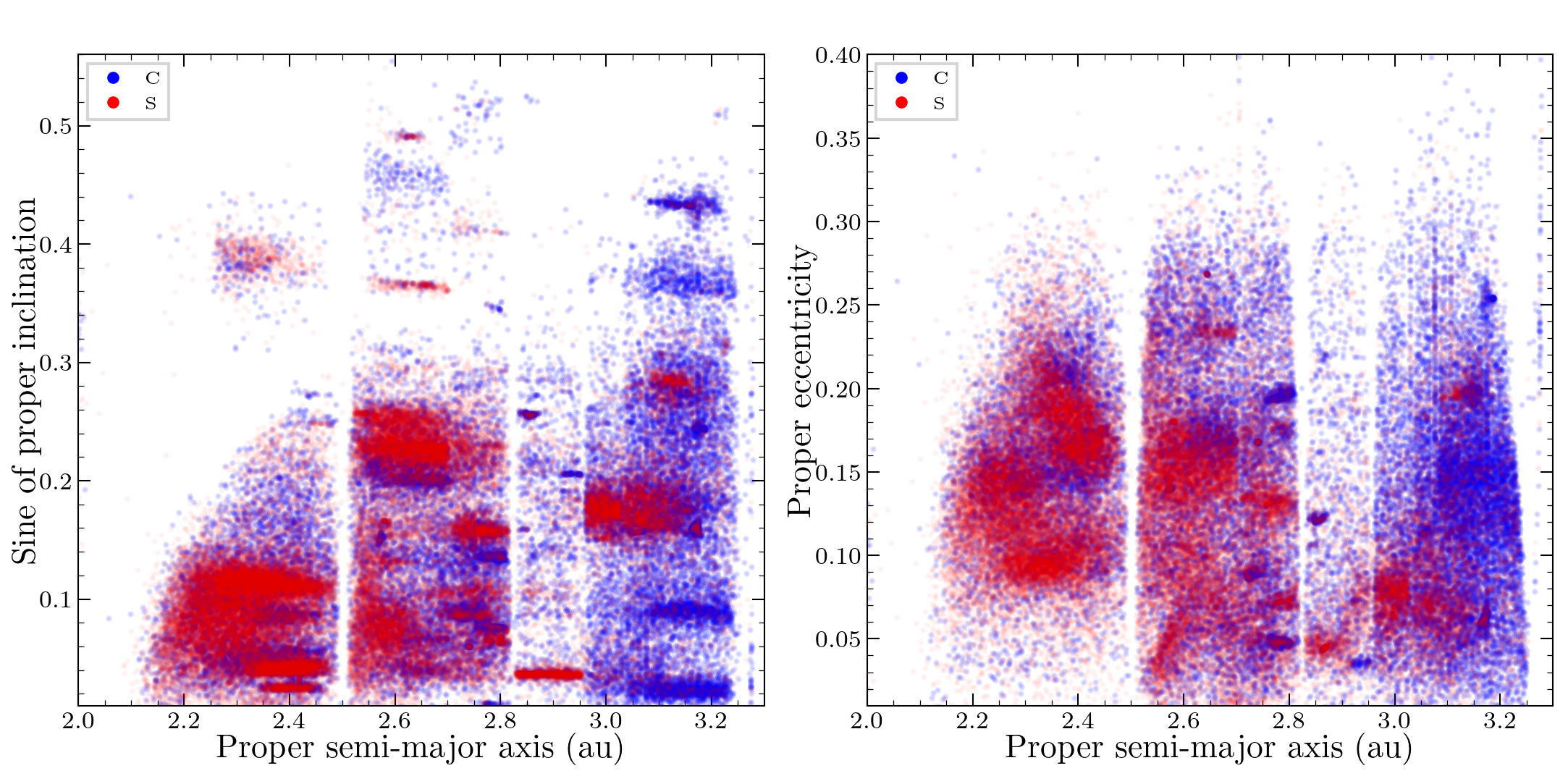}
  \caption{Orbital distribution of the \sm SSOs, color-coded by taxonomic class based on $g$-$r$ color only.}
  \label{fig:taxo_regions_gr}
\end{figure*}
 
 \begin{figure}[]
    \centering
    \input{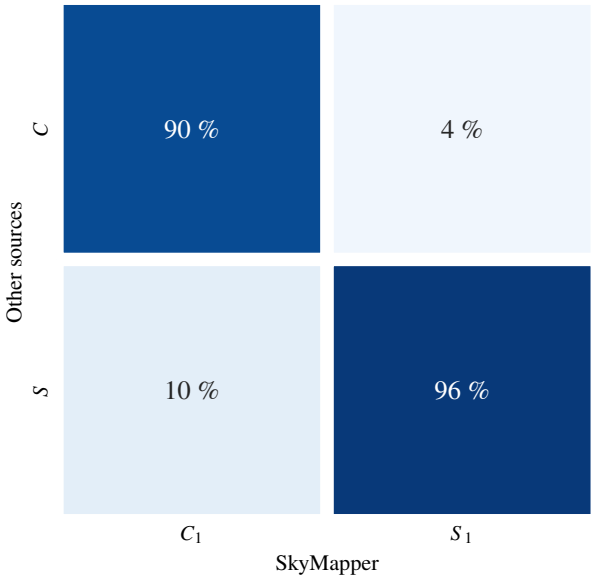}
    \caption{The confusion matrix of C and S taxonomy complexes of 1129 joint asteroids with known by spectra taxonomy and Skymapper taxonomy based on $g$-$r$ colors only.
    }
    \label{fig:matrix_gr}
\end{figure} 
  
  From these two normal distributions, we assign a probability of belonging
  to each complex to each asteroid. We then assign to each objects its most
  probable complex. Such way we classified \numb{71,822} asteroids as a $S_1$ type
  and \numb{40,590} as $C_1$ type,
  with the exception of the \numb{4,944} asteroids which probability did not differ by more than 
  \numb{10}\%.
  The distribution of $S_1$ and $C_1$ asteroids also reproduce the known distribution
  of compositions in the asteroids belt (\Autoref{fig:taxo_regions_gr}).

  It is interesting to note that even such a simplistic taxonomy criterion
  provides a very good agreement with spectral observations (\Autoref{fig:matrix_gr}).

\section{Future Work\label{sec:future}}

  The present catalog represents the first release of SSOs observed
  by the \sm Southern Survey, based on the data from its third data release (DR3).
  These data extend the multi-filter photometry provided by the 
  SDSS, which finished its imaging survey in 2009
  \citep[although still used by the community, see][for instance]{SergeyevCarry2021,2021Icar..36514494B}.
  \sm is still operating and we plan to release observations of SSOs in upcoming data 
  releases. \sm near-simultaneous acquisition of different filters provides instantaneous
  determination of colors, which can be used by the community 
  even with the start of operations of the upcoming
  Legacy Survey of Space and Time (LSST) by the Vera C. Rubin observatory
  that will require years to build the phase functions needed to determine
  SSO colors
  \citep{2009-EMP-105-Jones, 2021Icar..35414094M}.
  
  \sm multi-filter observations are rather unique as of today. 
  The observatories that submitted most data to the Minor Planet Center
  so far this year (2021) are
  Catalina Sky Survey,
  ATLAS (Asteroid Terrestrial-impact Last Alert System), and
  Pan-STARRS.
  While Catalina observed the most, it has a wide optical filter only.
  ATLAS uses two filters and its observing cadence allows for phase functions
  to be constructed \citep{2021Icar..35414094M},
  and a single color can be used for rough taxonomic classification
  \citep[see \Autoref{ssec:gr} and][]{2020ApJS..247...13E}.
  Pan-STARRS has lately contributed to most NEA discoveries, but has relied on 
  $g$, $r$, $i$ and a wide $w$ filters, precluding spectral classification
  as the $z$ filter probes the 1\,$\mu$m band
  (the most characteristic spectral feature in all major taxonomies since \citet{1975-Icarus-25-Chapman}.
  
  Finally, considering the limiting magnitude of \sm, it is unlikely that a large number
  of unknown SSOs may be discovered in its images, and we plan to proceed with the
  extraction of known
  SSOs only in future releases.

  \begin{figure}[t]
    \centering
    \input{figs/skm_sdss_matrix.pgf}
    \caption{The confusion matrix of \numb{6,965} asteroids taxonomy from the SDSS and \sm with the probability accuracy of more than 0.2}
    \label{fig:skm_sdss_matrix}
\end{figure}

\section{Conclusion}\label{sec:conclusion}

\begin{figure*}[t]
    \centering
    \input{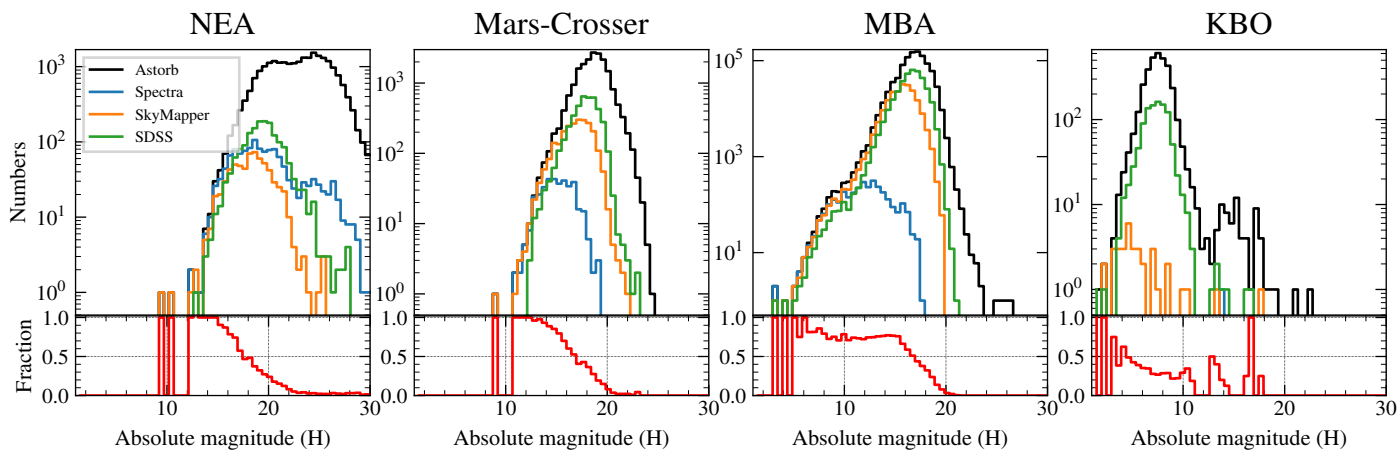}
    \caption{Comparison of the number
    of SSOs with either spectra, SDSS colors, or \sm colors,
    compared with the
    total number of SSO (taken from \texttt{astorb})
    for each of the NEA, Mars-Crossers, MBA, and KBO
    dynamical populations.
    }
    \label{fig:fractionH}
\end{figure*}

  \indent We extracted known Solar System objects from \sm Southern Survey
  DR3 images. 
  We applied a suite of filters to minimize contamination.
  We release a catalog of
  \numb{\skmtot} individual observations, consisting of
  \numb{\skmuniq} unique, known SSOs.
  The catalog contains the 
  \sm identification,
  astrometry,
  photometry, 
  SSO identification,
  geometry of observation, 
  and taxonomy.
  Its content is fully described in \Autoref{app:cat}.

  \indent The catalog completeness is estimated to 
  be about \numb{97}\% down to V\,=\,18~mag
  and the purity to almost
  \numb{100}\% down to V\,=\,20\,mag.
  The present catalog contains photometry of 
  \numb{669} near-Earth asteroids,
  \numb{2,487} Mars-crossers,
  \numb{196,269} main-belt asteroids, and
  \numb{1,928} Jupiter trojans.
  
  There are \numb{100,005} asteroids which were already observed by the SDSS
  \citep{SergeyevCarry2021}. 
  Combined, the SDSS and \sm data sets account almost 80\% of all MBAs with an
  absolute magnitude (H) between 10 and 16, each data set contributing equally.
  As most of the
  brighter asteroids (H$<$10) have spectra, the present \sm release represents a 
  major step toward completeness of taxonomic classification of asteroids in the main belt
  (\Autoref{fig:fractionH}).
  The situation is somewhat similar for NEAs and Mars-Crossers, with a completeness
  above 80\%
  of these two populations down to an absolute magnitude of 16, although spectral
  surveys contributed more
  \citep[e.g.,][]{2018PSS..157...82P, 2019AJ....158..196D, 2019Icar..324...41B}.
  The distant KBOs are the population with the lowest completeness of color characterization,
  below 50\% (caveat: we have not attempted to compile spectra of KBOs here).
  As \sm is an on-going survey, this completeness will increasing with future releases.
  
  We compared taxonomy classification of \numb{6,965} common asteroids between the
  recent SDSS release \citep{SergeyevCarry2021} and the present
  \sm catalog, considering only SSOs with a taxomony based on three colors
  and a probability above 0.2.
  The confusion matrix presented in \Autoref{fig:skm_sdss_matrix}
  shows a good
  agreement for the most common complexes like S, C, B, and V.
  Some confusion is present between less prominent classes, in particular between
  K/L/A and S.
  This highlights both the strength and the limitation of taxonomic classification
  based on broad-band colors only.

\begin{acknowledgements}
  This  research has been conducted within the NEOROCKS project, which
  has received funding from the European Union's Horizon 2020 research
  and innovation programme under grant agreement No 870403.
  CAO was supported by the Australian Research Council (ARC) through Discovery Project DP190100252. SWC acknowledges support from the National Research Foundation of
  Korea (NRF) grant, No.2020R1A2C3011091, funded by the Korea government (MSIT).\\

  \indent We thank J. Berthier, F. Spoto, M. Mahlke for discussions related to the present article.\\
 
  \indent The national facility capability for SkyMapper has been funded through ARC LIEF
  grant LE130100104 from the Australian Research Council, awarded to the University of Sydney,
  the Australian National University, Swinburne University of Technology,
  the University of Queensland, the University of Western Australia,
  the University of Melbourne, Curtin University of Technology, Monash University
  and the Australian Astronomical Observatory. SkyMapper is owned and operated by
  The Australian National University's Research School of Astronomy and Astrophysics.
  The survey data were processed and provided by the SkyMapper Team at ANU.
  The SkyMapper node of the All-Sky Virtual Observatory (ASVO) is hosted at the
  National Computational Infrastructure (NCI). Development and support of the
  SkyMapper node of the ASVO has been funded in part by Astronomy Australia Limited (AAL)
  and the Australian Government through the Commonwealth's Education Investment Fund (EIF)
  and National Collaborative Research Infrastructure Strategy (NCRIS),
  particularly the National eResearch Collaboration Tools and Resources (NeCTAR)
  and the Australian National Data Service Projects (ANDS).\\
 
  \indent The Pan-STARRS1 Surveys (PS1) and the PS1 public science archive have been
  made possible through contributions by the Institute for Astronomy, the
  University of Hawaii, the Pan-STARRS Project Office, the Max-Planck Society
  and its participating institutes, the Max Planck Institute for Astronomy,
  Heidelberg and the Max Planck Institute for Extraterrestrial Physics,
  Garching, The Johns Hopkins University, Durham University, the University of
  Edinburgh, the Queen's University Belfast, the Harvard-Smithsonian Center
  for Astrophysics, the Las Cumbres Observatory Global Telescope Network
  Incorporated, the National Central University of Taiwan, the Space Telescope
  Science Institute, the National Aeronautics and Space Administration under
  Grant No. NNX08AR22G issued through the Planetary Science Division of the
  NASA Science Mission Directorate, the National Science Foundation Grant No.
  AST-1238877, the University of Maryland, Eotvos Lorand University (ELTE),
  the Los Alamos National Laboratory, and the Gordon and Betty Moore
  Foundation.\\

  \indent This research made use of the cross-match service provided by CDS, Strasbourg
  \citep{2017AA...597A..89P},
  the IMCCE's \skybot and Skybot3D VO tools
  \citep{2006-ASPC-351-Berthier, 2016-MNRAS-458-Berthier},
  the JPL Horizons system \citep{1996DPS....28.2504G},
  the SVO Filter Profile Service (\url{http://svo2.cab.inta-csic.es/theory/fps/})
  supported from the Spanish MINECO through grant AYA2017-84089
  \citep{2012-IVOA-Rodrigo},
  and TOPCAT/STILTS
  \citep{2005ASPC..347...29T}.
  Thanks to the developers.
\end{acknowledgements}

\bibliographystyle{aa} 
\bibliography{current} 

\begin{appendix}

\section{Description of catalogs\label{app:cat}}

  We describe here the three catalogs of Solar System objects (SSOs) we release.
  The detection catalog (\Autoref{tab:cat_phot}) contains all the information on each observation.
  (mid-observing time, coordinates, etc.).
  The color catalogs contains all the measured SSO colors (\Autoref{tab:cat_obs_color}),
  while the object catalog (\Autoref{tab:cat_color_full}) contains a single entry per SSOs, with its average colors. 
  
  The \Autoref{tab:cat_color_ind} contains estimated taxonomy and orbital elements of asteroids. The most probably taxonomy depends of the color  priority in the follow sequence: $g-r$, $g-i$, $i-z$ colors have the first priority, two colors $g-r$, $i-z$, $g-i$, $i-z$, $r-i$, $i-z$ have a priority 2, 3, 4 consequently and $g-r$ color priority 5.
  

\begin{table}[h]
    \centering
    \caption{Description of the \sm catalog of SSO observations.}
    \label{tab:cat_phot}
    \begin{tabular}{l m{18mm} c m{4cm}}
    \hline \hline
    \textbf{ID} &\textbf{Name} & \textbf{Unit} & \textbf{Description} \\ 
    \hline
     1 & source & & Source unique identifier \\ 
     2 & frame & & Image unique identifier  \\ 
     3 & JD & day & Julian Date of observation \\ 
     4 & filter & & Filter name ($u,v,g,r,i,z$) \\ 
     5 & exptime & s & Exposure time \\ 
     6 & ra & deg & J2000 Right Ascension \\ 
     7 & dec & deg & J2000 Declination \\ 
     8 & psfMag & mag & PSF magnitude\\ 
     9 & psfMagErr & mag & PSF magnitude uncertainty \\ 
    10 & petroMag & mag & Petrosian magnitude\\ 
    11 & petroMagErr & mag & Petrosian magnitude uncertainty \\ 
    12 & number & & SSO IAU number \\ 
    13 & name & & SSO IAU designation \\ 
    14 & dynclass & & SSO dynamical class \\ 
    15 & ra\_rate & \arcsec/h & RA$\cos($DEC$)$ rate of motion \\ 
    16 & dec\_rate & \arcsec/h & DEC rate \\ 
    17 & V & mag &  Predicted visual magnitude \\ 
    \hline
\end{tabular}
\end{table}

\begin{table}[h]
    \centering
    \caption{Description of the color catalog extracted from all \sm SSOs observations.}
    \label{tab:cat_obs_color}
\begin{tabular}{l m{15mm} c m{4.3cm}}
    \hline\hline
    \textbf{ID} &\textbf{Name} & \textbf{Unit} & \textbf{Description} \\ 
    \hline 
    1 & number & & SSO IAU number \\ 
    2 & name & & SSO IAU designation \\ 
    3 & JD & day & Average epoch of observation \\ 
    4 & color & & Name of color (e.g., $g-r$) \\ 
    5 & dmag & mag & Value of the color \\ 
    6 & edmag & mag & Uncertainty on the color \\ 
    7 & dmjd & day & Delay between filters \\ 
    \hline
\end{tabular}
\end{table}

\begin{table}[h]
    \centering
    \caption{Description of the catalog of measured weighted mean SSOs colors.}
    \label{tab:cat_color_full}

\begin{tabular}{l m{15mm} c m{4.3cm}}
    \hline\hline
    \textbf{ID} &\textbf{Name} & \textbf{Unit} & \textbf{Description} \\ 
    \hline 
    1 & number & & SSO IAU number \\ 
    2 & name & & SSO IAU designation \\ 
    4 & color & & Name of color (e.g., $g-r$) \\ 
    5 & wdmag & mag & Weighted value of the color \\ 
    6 & var & mag & Weighted uncertainty of the color measurements \\
    7 & n &  & Number of color measurements, 0 - computed\\
    6 & emag & mag & Mean value of the color uncertainties \\ 
    7 & dmjd & day & Mean time difference between color measurements \\ 
    \hline
\end{tabular}
\end{table}


\begin{table}[h]
    \centering
    \caption{Description of individual \sm SSOs color catalog (measured and computed) and their taxonomy.}
    \label{tab:cat_color_ind}
    \begin{tabular}{l m{18mm} c m{4cm}}
        \hline\hline
        \textbf{ID} &\textbf{Name} & \textbf{Unit} & \textbf{Description} \\ 
        \hline 
        1 & number & & SSO IAU number \\ 
        2 & name & & SSO IAU designation \\ 
        3 & dynclass & & SSO dynamic class \\ 
        4-18 & m[$^*$color~list] & mag & set of the color magnitude values \\
        29-43 & e[$^*$color~list] & mag & set of the color uncertainties \\
        44-58 & d[$^*$color~list] & day & set of time values between observations \\
        
        59-69 & p[C, S, B, L, D, K, X, Q, V, A, U] & & probability of the complex values \\ 
    70 & complex & & Most probably complex \\
    71 & pcomplex & & Probability value of the complex \\
    72 & complex1 & & First the most probably complex \\
    73 & complex2 & & Second the most probably complex \\
    74 & pcomplex1 & & Probability value of first the most probably complex \\
    75 & pcomplex2 & & Probability value of second the most probably complex \\
    
    76 & nc & & Number of colors where used for the taxonomy estimation \\
    
%
        \hline
\end{tabular}
    \text{$^*$ color list: u-v,u-g,u-r,u-i,u-z,v-g,v-r,v-i,v-z,g-r,g-i,i-z,g-z,r-i,r-z}
\end{table}

\newpage
\newpage
\section{Multi-color taxonomy}

We present here the color distribution, pseudo-reflectance, confusion matrix, orbital distribution and taxonomy boundaries of asteroids with two colors. 

 \begin{figure*}[]
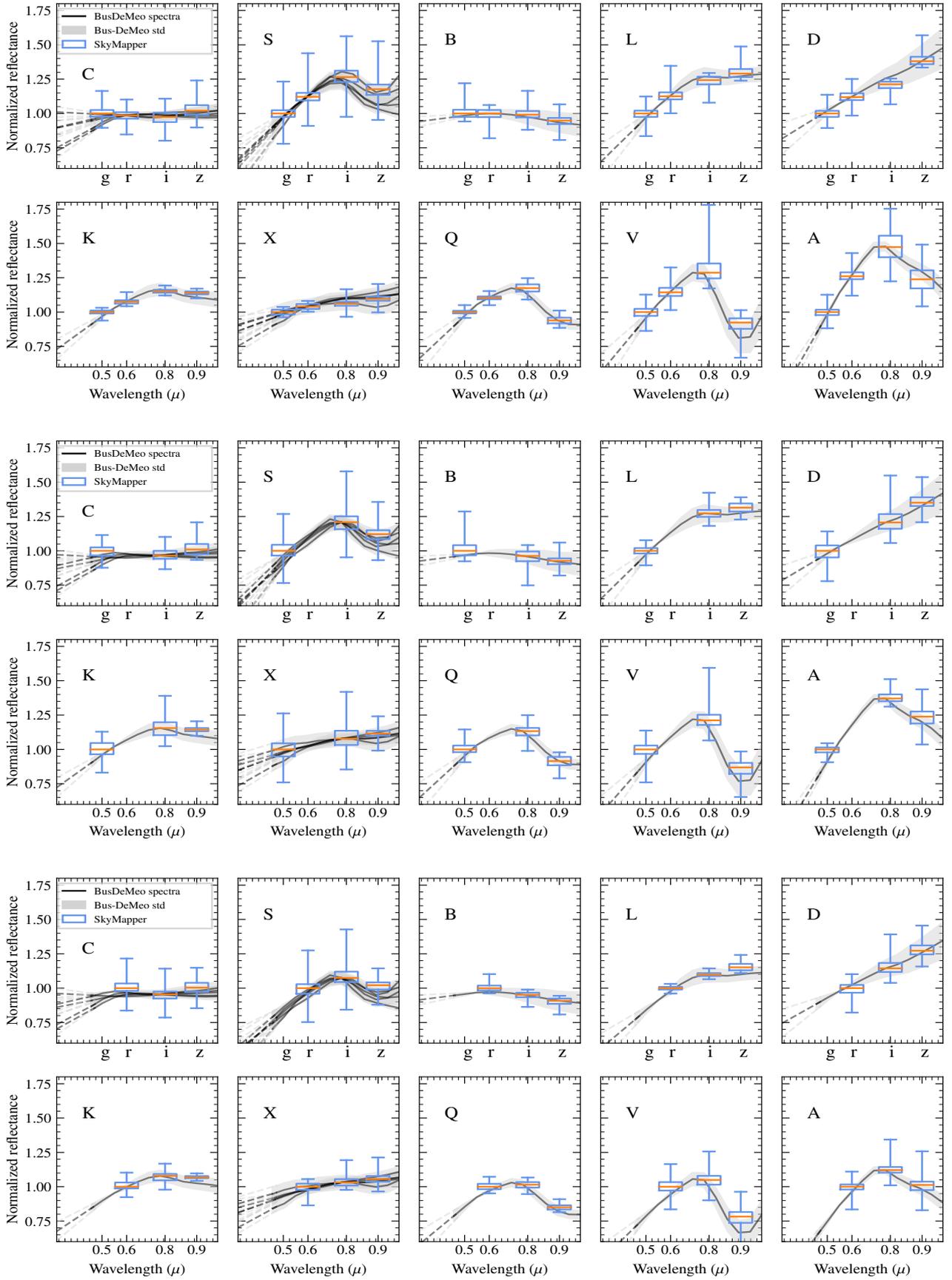

    \centering
    \scalebox{0.95}[0.85]{\input{figs/gr_iz_classes.pgf}}
    \scalebox{0.95}[0.85]{\input{figs/gi_iz_classes.pgf}}
    \scalebox{0.95}[0.85]{\input{figs/ri_iz_classes.pgf}}
    \caption{Grouped by taxonomic class pseudo-reflectance spectra of asteroids 
    based on two color taxonomy (top: $g-r$, $i-z$ colors), (middle: $g-i$, $i-z$ colors) and (bottom: $r-i$, $i-z$ colors). We indicate the average wavelength of each filter in the lower plots.
        The distribution of values for each band is
        represented by whiskers (95\% extrema, and the 25, 50, and 75\% quartiles).
        For each, we also represent
        the associated template spectra of the Bus-DeMeo taxonomy
        \citep{2009Icar..202..160D}.}
    \label{fig:twocolor_reflectance}
\end{figure*}

\begin{figure}[]
    \centering
    \scalebox{0.85}[0.85]{\includegraphics{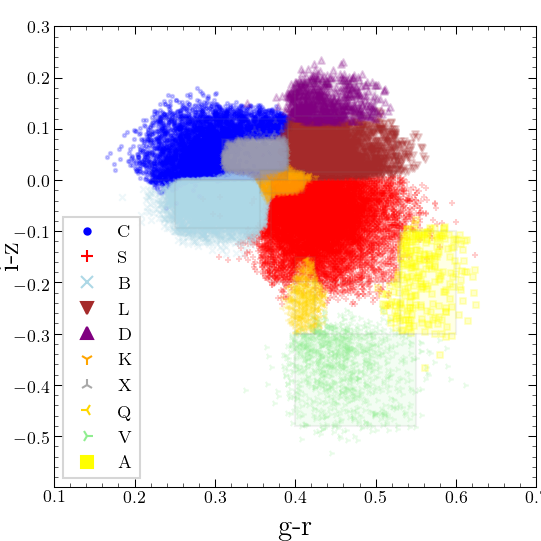}}
    \scalebox{0.85}[0.85]{\includegraphics{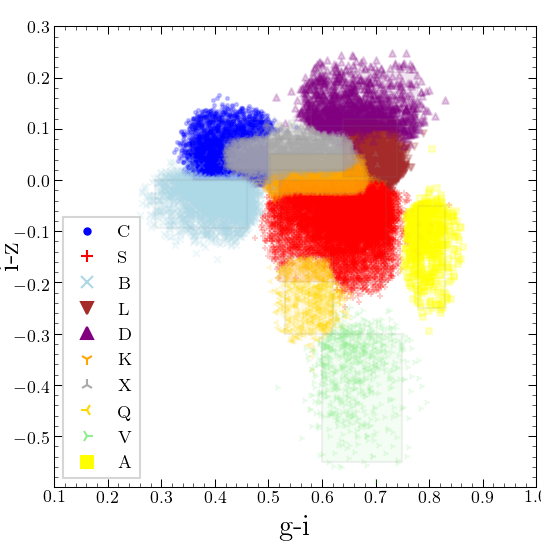}}
    \scalebox{0.85}[0.85]{\includegraphics{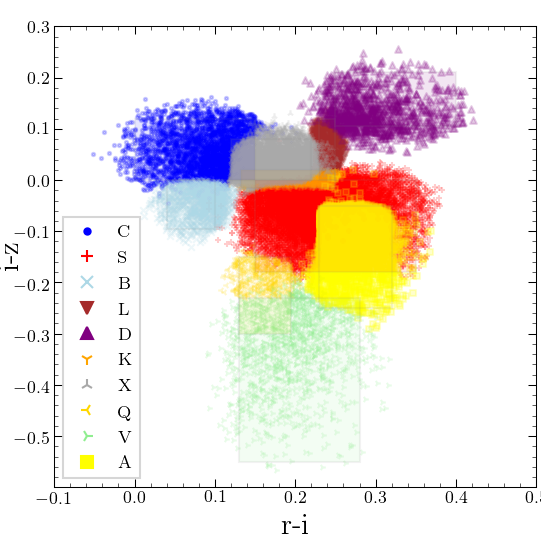}}
    \caption{Taxonomy of \sm asteroids with a probability value more than \numb{0.2} with two (top: $g-r$, $i-z$, middle: $g-i$, $i-z$ and bottom: $r-i$, $i-z$) colors.
    Boxes show the boundaries of the taxonomic classes. Color points mark an individual asteroids.}
    \label{fig:griz_taxonomy}
\end{figure}

\begin{figure}[]
    \centering
    \scalebox{0.85}[0.85]{\input{figs/griz_matrix.pgf}}
    \scalebox{0.85}[0.85]{\input{figs/giiz_matrix.pgf}}
    \scalebox{0.85}[0.85]{\input{figs/riiz_matrix.pgf}}
    \caption{Confusion matrices of asteroids taxonomy based on (top: $g-r$, $i-z$, middle: $g-i$, $i-z$ and bottom: $r-i$, $i-z$) \sm colors opposite taxonomy from published spectral observations. The values are reported in percents.}
    \label{fig:matrix_griz}
\end{figure}

\begin{figure*}[]
  \centering
  \scalebox{0.85}[0.85]{\includegraphics{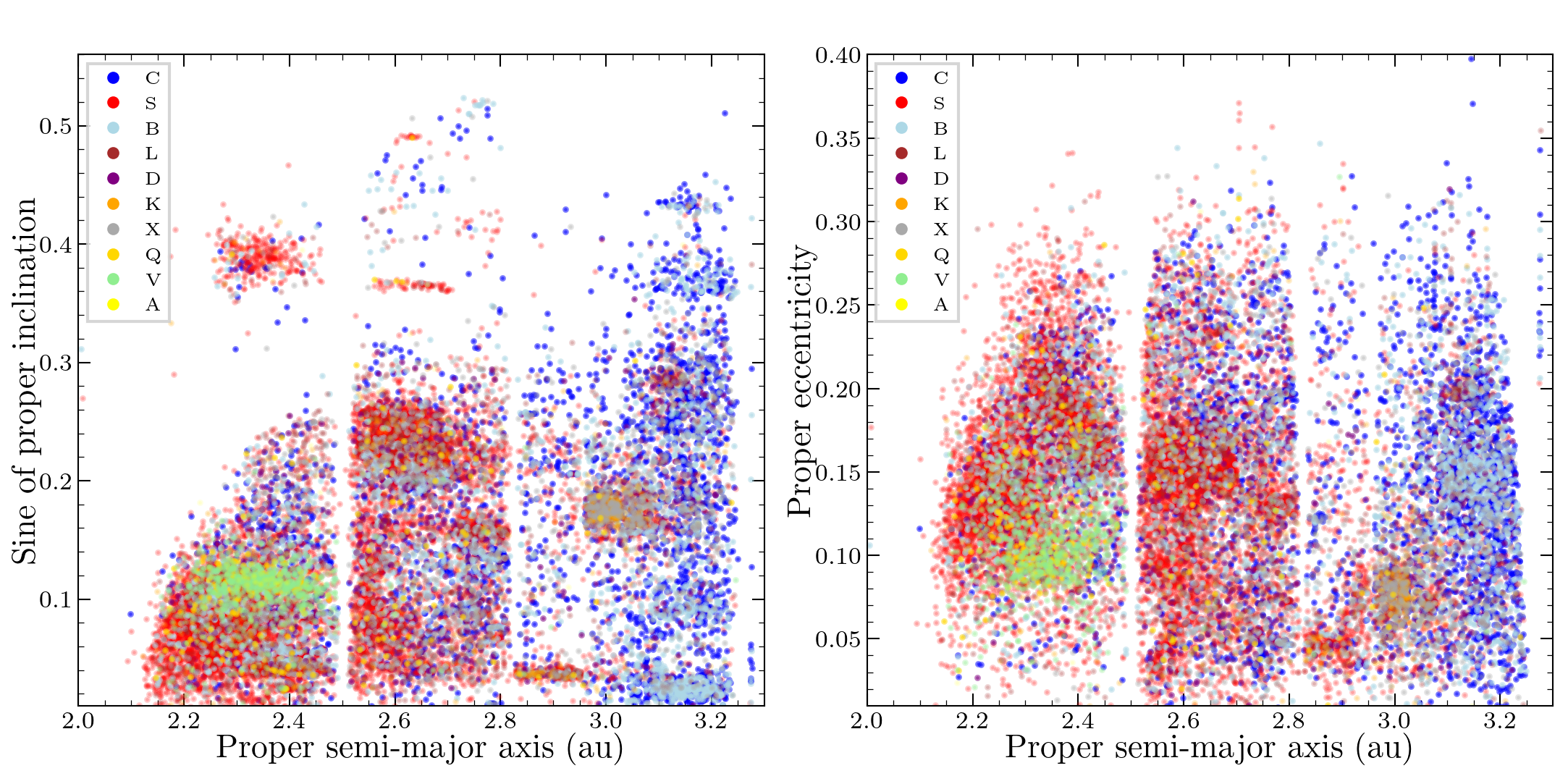}}
  \scalebox{0.85}[0.85]{\includegraphics{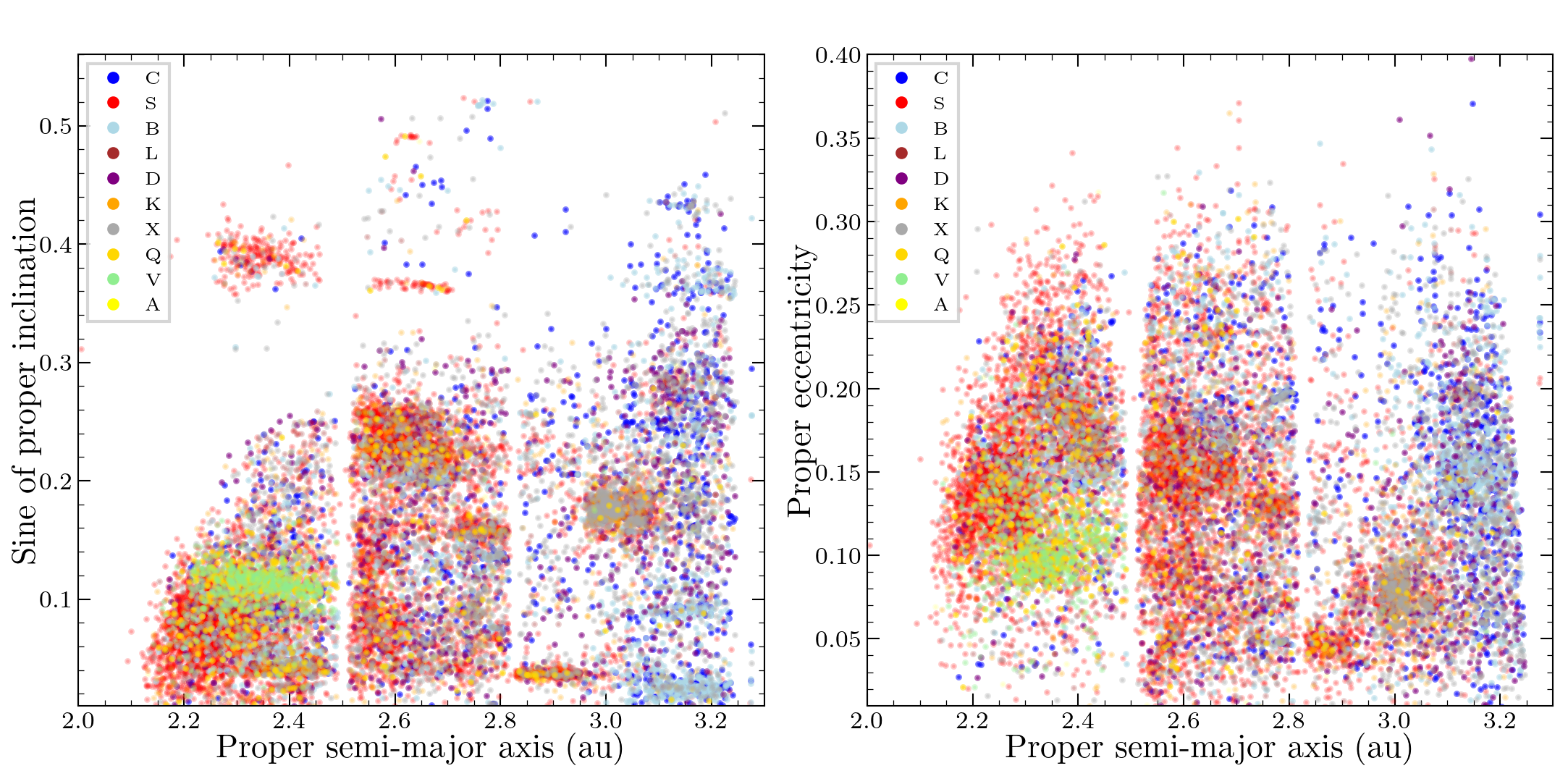}}
  \scalebox{0.85}[0.85]{\includegraphics{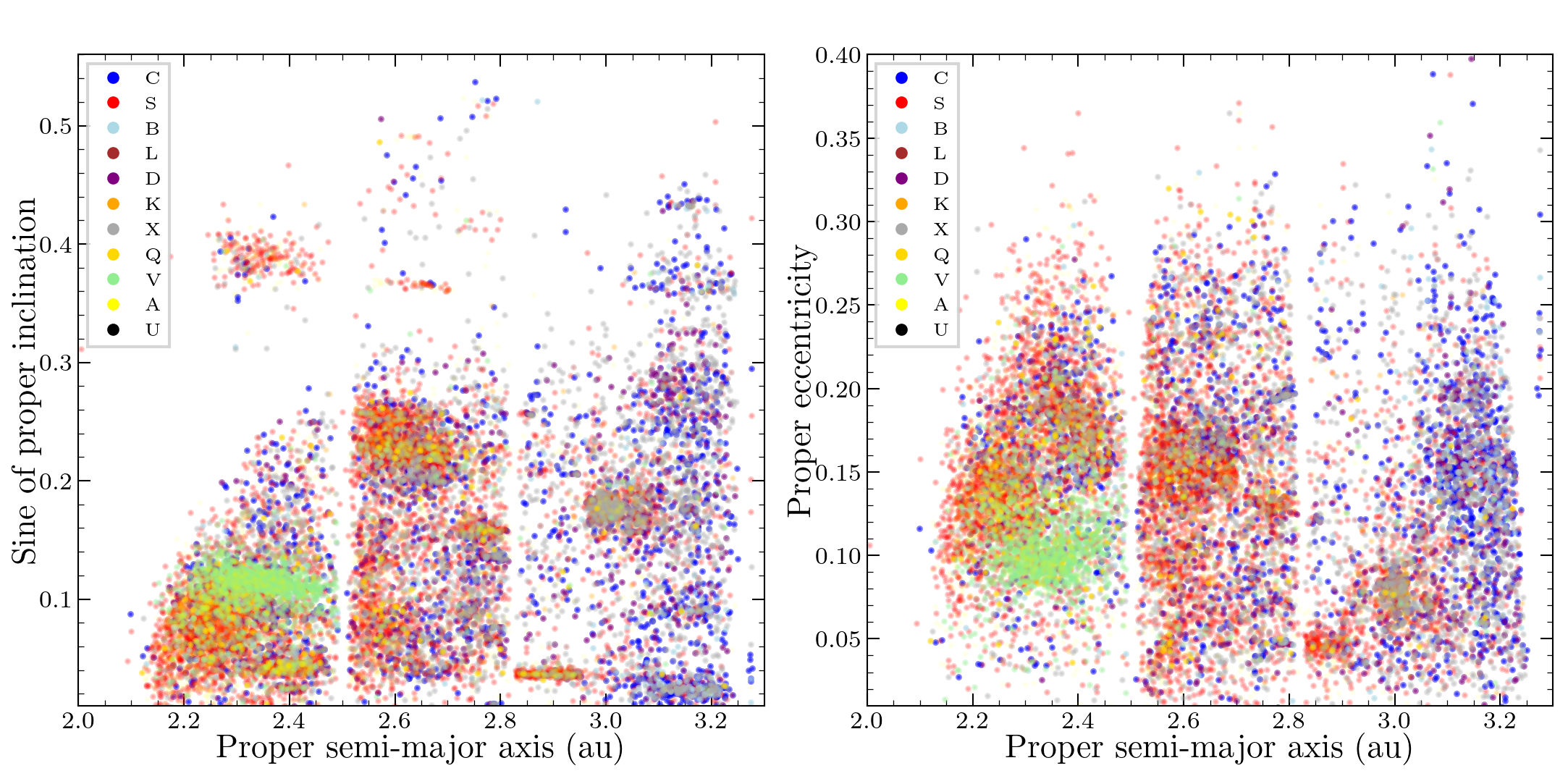}}
  \caption{Orbital distribution of the \sm SSOs, color-coded by taxonomic class based on their (top: $g-r$, $i-z$, middle: $g-i$, $i-z$ and bottom: $r-i$, $i-z$) colors}
  \label{fig:taxo_regions_griz}
\end{figure*}

\begin{table}[]
    \centering
    \caption{Boundaries of taxonomy complexes in g-r, g-i, i-z color space.}
    \label{tab:taxo_bound_grgiiz}    
    \begin{tabular}{ccccccc}
    \hline
    \hline
    complex & $gr_{min}$ & $gr_{max}$ & $iz_{min}$ & $iz_{max}$ & $gi_{min}$ & $gi_{max}$ \\
    \hline
    C &   0.220 &   0.350 &   0.000 &   0.120 &   0.360 &    0.50 \\
    S &   0.370 &   0.530 &  -0.200 &   0.000 &   0.620 &    0.73 \\
    B &   0.250 &   0.356 &  -0.094 &   0.001 &   0.288 &    0.46 \\
    L &   0.391 &   0.527 &   0.015 &   0.120 &   0.650 &    0.78 \\
    D &   0.400 &   0.500 &   0.105 &   0.210 &   0.650 &    0.80 \\
    K &   0.370 &   0.430 &  -0.015 &   0.020 &   0.503 &    0.68 \\
    X &   0.280 &   0.400 &   0.000 &   0.080 &   0.400 &    0.62 \\
    Q &   0.400 &   0.425 &  -0.300 &  -0.150 &   0.530 &    0.62 \\
    V &   0.350 &   0.600 &  -0.550 &  -0.230 &   0.520 &    0.75 \\
    A &   0.459 &   0.595 &  -0.250 &  -0.050 &   0.780 &    0.83 \\
\hline
\end{tabular}

\end{table}

\begin{table}[]
    \centering
    \caption{Boundaries of taxonomy complexes in g-r, i-z color space.}
    \label{tab:taxo_bound_griz}    
    \begin{tabular}{ccccc}
    \hline
    \hline
    complex & $gr_{min}$ & $gr_{max}$ & $iz_{min}$ & $iz_{max}$ \\
    \hline
    C &   0.220 &   0.391 &   0.000 &   0.120 \\
    S &   0.370 &   0.530 &  -0.200 &   0.000 \\
    B &   0.250 &   0.356 &  -0.094 &   0.001 \\
    L &   0.391 &   0.527 &   0.015 &   0.100 \\
    D &   0.400 &   0.500 &   0.125 &   0.190 \\
    K &   0.356 &   0.411 &  -0.025 &   0.020 \\
    X &   0.309 &   0.391 &   0.020 &   0.080 \\
    Q &   0.400 &   0.425 &  -0.300 &  -0.150 \\
    V &   0.400 &   0.550 &  -0.480 &  -0.300 \\
    A &   0.530 &   0.600 &  -0.300 &  -0.100 \\
\hline
\end{tabular}

\end{table}

\begin{table}[]
    \centering
    \caption{Boundaries of taxonomy complexes in g-i, i-z color space.}
    \label{tab:taxo_bound_giiz}    
    \begin{tabular}{ccccc}
    \hline
    \hline
    complex & $gi_{min}$ & $gi_{max}$ & $iz_{min}$ & $iz_{max}$ \\
    \hline
    C &   0.360 &   0.500 &   0.000 &   0.120 \\
    S &   0.520 &   0.720 &  -0.200 &   0.000 \\
    B &   0.288 &   0.460 &  -0.094 &   0.001 \\
    L &   0.640 &   0.740 &   0.005 &   0.120 \\
    D &   0.580 &   0.780 &   0.105 &   0.210 \\
    K &   0.503 &   0.680 &  -0.025 &   0.050 \\
    X &   0.417 &   0.658 &   0.020 &   0.080 \\
    Q &   0.530 &   0.620 &  -0.300 &  -0.150 \\
    V &   0.600 &   0.750 &  -0.550 &  -0.300 \\
    A &   0.780 &   0.830 &  -0.250 &  -0.050 \\
    \hline
\end{tabular}

\end{table}

\begin{table}[]
    \centering
    \caption{Boundaries of taxonomy complexes in r-i, i-z color space.}
    \label{tab:taxo_bound_riiz}    
    \begin{tabular}{ccccc}
    \hline
    \hline
    complex & $ri_{min}$ & $ri_{max}$ & $iz_{min}$ & $iz_{max}$ \\
    \hline
    C &   0.000 &   0.150 &   0.000 &   0.120 \\
    S &   0.150 &   0.350 &  -0.180 &   0.000 \\
    B &   0.040 &   0.100 &  -0.095 &   0.001 \\
    L &   0.230 &   0.250 &   0.000 &   0.120 \\
    D &   0.250 &   0.400 &   0.105 &   0.210 \\
    K &   0.133 &   0.250 &  -0.015 &   0.020 \\
    X &   0.120 &   0.220 &   0.000 &   0.080 \\
    Q &   0.130 &   0.195 &  -0.300 &  -0.150 \\
    V &   0.130 &   0.280 &  -0.550 &  -0.230 \\
    A &   0.230 &   0.320 &  -0.250 &  -0.050 \\
\hline
\end{tabular}

\end{table}

\end{appendix}

\end{document}